\documentclass[draftcls,journal,onecolumn]{IEEEtran}
\usepackage{graphicx,tikz}
\usepackage{amsmath,amssymb}
\usepackage{algorithm,algorithmic}
\usepackage{cite}
\usetikzlibrary{shadows}

\newcommand{\emb}[1]{\tilde{#1}}
\newcommand{\proj}[1]{\bar{#1}}
\newcommand{\repr}[1]{\tilde{#1}}
\newcommand{\best}[1]{{\sigma}^*(#1)}
\newcommand{\mat}[1]{\begin{bmatrix} #1 \end{bmatrix}}
\newcommand{\realL}{\Lambda_r}
\newcommand{\realLc}{\Lambda'_r}
\newcommand{\complexL}{\Lambda}
\newcommand{\complexLc}{\Lambda'}
\newcommand{\nn}{\text{NN}}
\newcommand{\optt}{\textsf{opt}}
\newcommand{\hh}{^{\textsf{H}}}
\renewcommand{\Re}{\textrm{Re}}
\renewcommand{\Im}{\textrm{Im}}

\DeclareMathOperator{\diag}{diag}
\DeclareMathOperator{\gl}{GL}
\DeclareMathOperator{\SNR}{\textsf{SNR}}
\DeclareMathOperator{\Rmes}{\textsf{R}_{\textsf{mes}}}

\newcommand{\ZZ}{\mathbb{Z}}

\newcommand{\RR}{\mathbb{R}}
\newcommand{\CC}{\mathbb{C}}
\newcommand{\FF}{\mathbb{F}}

\newcommand{\calA}{\mathcal{A}}
\newcommand{\calB}{\mathcal{B}}
\newcommand{\calC}{\mathcal{C}}
\newcommand{\calD}{\mathcal{D}}
\newcommand{\calE}{\mathcal{E}}

\newcommand{\calH}{\mathcal{H}}

\newcommand{\calL}{\mathcal{L}}

\newcommand{\calQ}{\mathcal{Q}}
\newcommand{\calR}{\mathcal{R}}
\newcommand{\calS}{\mathcal{S}}

\newcommand{\calV}{\mathcal{V}}

\newcommand{\bfA}{\mathbf{A}}
\newcommand{\bfB}{\mathbf{B}}

\newcommand{\bfD}{\mathbf{D}}

\newcommand{\bfG}{\mathbf{G}}

\newcommand{\bfI}{\mathbf{I}}
\newcommand{\bfJ}{\mathbf{J}}

\newcommand{\bfL}{\mathbf{L}}
\newcommand{\bfM}{\mathbf{M}}

\newcommand{\bfP}{\mathbf{P}}
\newcommand{\bfQ}{\mathbf{Q}}

\newcommand{\bfU}{\mathbf{U}}
\newcommand{\bfV}{\mathbf{V}}
\newcommand{\bfW}{\mathbf{W}}

\newcommand{\bfY}{\mathbf{Y}}

\newcommand{\bfa}{\mathbf{a}}
\newcommand{\bfb}{\mathbf{b}}
\newcommand{\bfc}{\mathbf{c}}
\newcommand{\bfd}{\mathbf{d}}

\newcommand{\bfg}{\mathbf{g}}
\newcommand{\bfh}{\mathbf{h}}

\newcommand{\bfn}{\mathbf{n}}

\newcommand{\bfr}{\mathbf{r}}

\newcommand{\bfu}{\mathbf{u}}
\newcommand{\bfv}{\mathbf{v}}
\newcommand{\bfw}{\mathbf{w}}
\newcommand{\bfx}{\mathbf{x}}
\newcommand{\bfy}{\mathbf{y}}
\newcommand{\bfz}{\mathbf{z}}
\newcommand{\bfzero}{\mathbf{0}}
\newcommand{\bflambda}{\boldsymbol{\lambda}}

\newtheorem{thm}{Theorem}
\newtheorem{prop}{Proposition}

\newtheorem{lem}{Lemma}
\newtheorem{defi}{Definition}
\newtheorem{exam}{Example}

\title{An Algebraic Approach to Physical-Layer\\ Network Coding}

\author{Chen~Feng,
Danilo~Silva,~\IEEEmembership{Member,~IEEE}, and
Frank~R.~Kschischang,~\IEEEmembership{Fellow,~IEEE}%
\thanks{Manuscript received July 21, 2011; revised Sep. 17, 2012 and
July 3, 2013. The work of D. Silva was supported in part by the S\~ao Paulo Research Foundation (FAPESP-Brasil) and in part by the Brazilian National Research Council (CNPq) under Grant 482131/2010-1.
This paper was presented in part at the IEEE International
Symposium on Information Theory, Austin, TX, June 2010, and in part at
the Annual Conference on Information Sciences and Systems,
Baltimore, MD, March 2011.}%
\thanks{C.~Feng and F.~R.~Kschischang are with the Department of
Electrical and Computer Engineering, University of Toronto, Canada (email: \texttt{cfeng@eecg.utoronto.ca}; \texttt{frank@comm.utoronto.ca}).}
\thanks{D.~Silva is with the Department of Electrical Engineering, Federal University of Santa Catarina, Brazil (email: \texttt{danilo@eel.ufsc.br}).}%
}
\IEEEpubid{0000--0000/00\$00.00~\copyright~2013 IEEE}
	
\begin{document}

\maketitle

\begin{abstract}
The problem of designing physical-layer network coding (PNC) schemes via
nested lattices is considered. Building on the compute-and-forward
(C\&F) relaying strategy of Nazer and Gastpar, who demonstrated its
asymptotic gain using information-theoretic tools, an algebraic approach
is taken to show its potential in practical, non-asymptotic, settings. A
general framework is developed for studying nested-lattice-based PNC
schemes---called lattice network coding (LNC) schemes for short---by
making a direct connection between C\&F and module theory. In
particular, a generic LNC scheme is presented that makes no assumptions
on the underlying nested lattice code. C\&F is re-interpreted in this
framework, and several generalized constructions of LNC schemes are
given. The generic LNC scheme naturally leads to a linear network coding
channel over modules, based on which non-coherent network coding can be
achieved. Next, performance/complexity tradeoffs of LNC schemes are
studied, with a particular focus on hypercube-shaped LNC schemes. The
error probability of this class of LNC schemes is largely determined by
the minimum inter-coset distances of the underlying nested lattice code.
Several illustrative hypercube-shaped LNC schemes are designed based on
Construction~A and D, showing that nominal coding gains of $3$ to
$7.5$~dB can be obtained with reasonable decoding complexity. Finally,
the possibility of decoding multiple linear combinations is considered
and related to the shortest independent vectors problem. A notion of
dominant solutions is developed together with a suitable
lattice-reduction-based algorithm.
\end{abstract}


\section{Introduction}
\label{sec:intro}

\IEEEPARstart{N}{ested}\scalebox{0.94}[1.0]{-lattice-based physical-layer network coding} \scalebox{0.94}[1.0]{(LNC)} is a type of compute-and-forward (C\&F) relaying strategy
\cite{NG09} that is emerging as a compelling information transmission
scheme in Gaussian relay networks. LNC exploits the property that
integer linear combinations of lattice points are again lattice points.
Based on this property, relays in LNC attempt to decode their received
signals into integer linear combinations of codewords, which they then
forward. This approach induces an end-to-end network coding channel from
which the transmitted information can be recovered by solving a linear
system.

In this paper, we develop a generic LNC scheme that makes no particular
assumption on the structure of the underlying nested lattice code,
thereby enabling a variety of code-design techniques. A key aspect of
this approach is a so-called ``linear labeling'' of the points in a
nested lattice code that gives rise to a beneficial compatibility
between the $\CC$-linear arithmetic operations performed by the wireless
channel and the linear operations in the message space that are required
for linear network coding. Similar to vector-space-based noncoherent
network coding (e.g., \cite{KK07}), the linear labelings of this paper
induce a noncoherent end-to-end network coding channel with a message
space having, in general, a module-theoretic algebraic structure,
thereby providing a foundation for achieving noncoherent network coding
over general Gaussian relay networks.

We study the error performance of a class of hypercube-shaped LNC
schemes, and show that the error performance is largely determined by
the minimum inter-coset distance of the underlying nested lattice code.
By way of illustration, we adapt several known lattice constructions to
give three exemplar LNC schemes that provide nominal coding gains of 3
to 7.5~dB while admitting reasonable decoding complexity.

We also study the possibility that a relay may attempt to decode more
than one linearly independent combination of messages, and we relate
this problem to the ``shortest independent vectors problem'' in lattices
\cite{Blomer00}. For this problem, a notion of dominant solutions is
introduced together with a lattice-reduction-based algorithm, which may
be of independent interest.

\IEEEpubidadjcol

LNC can be seen as generalization of several previous physical layer
network coding (PNC) schemes \cite{ZLL06, PY06, NG06}. The earliest PNC
schemes were applied to a two-way relay channel in which the relay
attempts to decode the modulo-two sum (XOR) of the transmitted messages.
It was observed in \cite{PY07, APT08} that the XOR can be replaced by a
family of functions satisfying the so-called ``exclusive law of network
coding.'' Furthermore, the choice of function can potentially be adapted
to the instantaneous channel realizations, although a complicated
computer search may be needed \cite{APT08} to choose the function
optimally, even in the case of low-dimensional constellations such as
$16$-QAM. Because LNC considers only linear combinations, not general
functions, it provides an efficient method, even in high-dimensional
spaces, to perform such channel-adaptive decoding. Further PNC schemes
presented in \cite{NWS07, NCL08, WN09, AST10} aim to approach the
capacities of various two-way relay channels. A survey of PNC for
two-way relay channels can be found in \cite{PK09}.

\begin{figure}[t]
\centering
\includegraphics[width=0.33\textwidth]{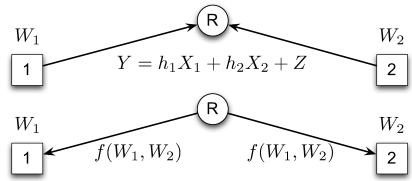}
\caption{Illustration of a two-round physical-layer network coding scheme.}
\label{fig:twrc}
\end{figure}

The use of nested lattice codes (or Voronoi constellations) in PNC was
first proposed in \cite{NG06, NWS07}, leading to the development of C\&F
relaying. A key feature of the C\&F strategy is that no channel state
information (CSI) is required at the transmitters. In contrast to
alternative advanced strategies such as noisy network coding
\cite{LKGC10} and quantize-map-and-forward strategy \cite{ADT11, OD10},
the C\&F strategy does not require global channel-gain information at
the destinations. All of these make C\&F an appealing candidate for
practical implementation.

The C\&F strategy can be enhanced by assuming CSI at the transmitters
\cite{NW11} or by installing multiple antennas at the relays and
destinations \cite{ZNGE09,BC10}. Practical code constructions for C\&F
are presented (see, e.g., \cite{HN11, OZEGN11, FSK11_isit, B11}). A
recent survey of C\&F can be found in \cite{NG11}.

After the conference publication of an earlier version of this work
\cite{FSK10} (see also \cite{FSK11-ciss, FSK11-cwit}), several papers
have appeared following our algebraic framework. For example, the work
of \cite{QJ12} presents several design examples based on Eisenstein
lattices, which can achieve a shaping gain of 0.167 dB compared to our
examples based on Gaussian lattices. The work of \cite{TNBH12} studies
the existence of asymptotically-good nested lattices over Eisenstein
integers, which can offer higher computation rates for certain channel
realizations compared to the computation rates in \cite{NG09} (which are
based on Gaussian integers).

The remainder of this paper is organized as follows.
Section~\ref{sec:motivating} presents motivating examples to illustrate
the role of algebra in PNC. Section~\ref{sec:preli} reviews some
well-known mathematical preliminaries that will be useful in setting up
our algebraic framework. Section~\ref{sec:formulation} presents a
problem formulation of linear PNC and summarizes some of Nazer-Gastpar's
main results in the context of our formulation. Section~\ref{sec:lnc}
studies the algebraic properties of LNC, presenting a generic LNC scheme
that induces an end-to-end linear network coding channel over modules.
Section~\ref{sec:error_prob} turns to the geometric properties of LNC,
presenting a union bound estimate as well as some design criteria.
Section~\ref{sec:design} contains several illustrative design examples
for practical LNC schemes, showing that a decent nominal coding gain is
quite possible under practical constraints. Section~\ref{sec:choice}
studies the problem of choosing multiple coefficient vectors, which is
closely related to some known lattice problems.
Section~\ref{sec:simulation} presents simulation results, while
Section~\ref{sec:conclusion} concludes this paper.

\section{Motivating Examples}\label{sec:motivating}

In this section, we illustrate the role of algebra in PNC with a
particular focus on two-way relay channels, where two terminals attempt
to exchange their messages $W_1, W_2$ through a central relay, as shown
in Fig.~\ref{fig:twrc}. For this channel model, a PNC scheme consists of
two rounds of communication. In the first round, the terminals
simultaneously transmit their signals $X_1, X_2$ to the relay, and the
relay tries to decode a function $f(W_1,W_2)$ of the messages from the
received signal $Y$. In the second round, the relay broadcasts the
decoded function $f(W_1, W_2)$ to the terminals, based on which each
terminal recovers the other message with its own message held as side
information.

\begin{figure}[t]
\centering\scalebox{0.9}{\includegraphics{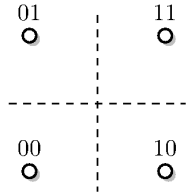}}
\caption{Transmitted QPSK constellation.}
\label{fig:QPSK}
\end{figure}

To illustrate how a PNC scheme works, we assume that the channels
between terminals and the relay are complex-valued flat-fading channels
with additive white Gaussian noise, that the messages $W_1,W_2$ take
values in the set $\{ 00, 01, 10, 11 \}$, and that (uncoded)
Gray-labeled quaternary phase-shift-keying (QPSK) modulation is used,
with the signal constellation given in Fig.~\ref{fig:QPSK}. The channel
gains between the terminals and the relay are denoted as $h_1$ and
$h_2$. Furthermore, we assume that the relay aims to decode the XOR of
the messages.

\begin{figure}[t]
\centering\begin{tabular}{c}
\includegraphics[scale=0.9]{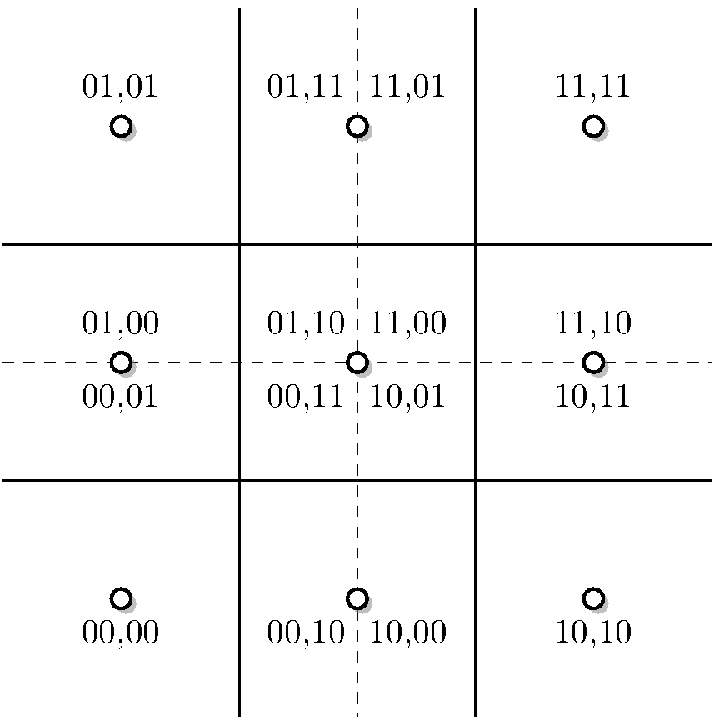}\\{\small (a)}\\
\includegraphics[scale=0.9]{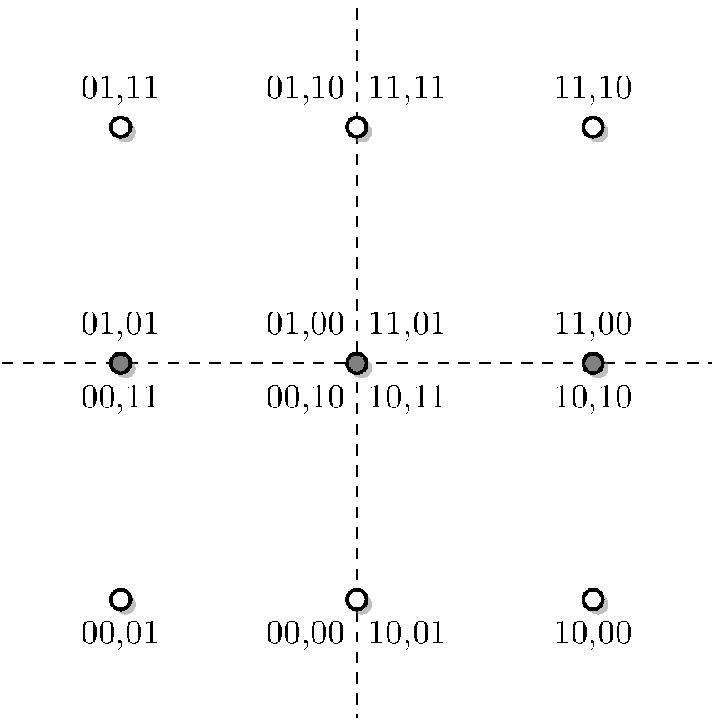}\\{\small (b)}
\end{tabular}
\caption{Received constellations with QPSK when (a) $h_1 = h_2 = 1$,
and (b) $h_1=1$, $h_2 = i$.}
\label{fig:qpsk11}
\end{figure}

We first consider the ideal special case in which the channel gains are
precisely unity, i.e., $h_1 = h_2 = 1$. The received constellation is
depicted in Fig.~\ref{fig:qpsk11}(a), together with the decision region
for XOR decoding. Although some received points are overlapping, say
point $(W_1,W_2)=(01, 11)$ and point $(11, 01)$, the overlapping points
have the same XOR value, resulting in no ambiguity.

Next, suppose that the channel gains are $h_1 = 1, h_2 = i$. In this
scenario, unfortunately, overlapping points have different XOR values;
see Fig.~\ref{fig:qpsk11}(b). For instance, point $(01, 10)$ has XOR
value $01 \oplus 10 = 11$; whereas point $(11, 11)$ has XOR value $00$.

To solve this ambiguity, one natural attempt is to let the relay decode
some linear function instead of the XOR. For example, if the relay
interprets each message $W_\ell = [w_{\ell 1}\ w_{\ell 2}]$ ($\ell = 1,
2$) as an element in $\FF_4$ by mapping it to $w_{\ell 1} \alpha +
w_{\ell 2}$ (where $\alpha$ is a primitive element of $\FF_4$) and tries
to decode the function $f_1(W_1, W_2) = W_1 + \alpha W_2$, then both
point $(01, 10)$ and point $(11, 11)$ give rise to the same value $10$.
However, there are still some ambiguities that cannot be resolved by
this function (the shaded dots in Fig.~\ref{fig:qpsk11}(b)).

In fact, no linear functions over $\FF_4$ can resolve all the
ambiguities in the received constellation, and the relay has to make use
of the structure of a finite ring rather than that of a finite field.
Specifically, let the relay interpret each message $W_\ell = [w_{\ell
1}\ w_{\ell 2}]$ as $w_{\ell 1} + w_{\ell 2} i \in \ZZ_2[i]$ with
addition and multiplication defined as
\begin{align*}
	a + bi + c + di & = [a + c]_2 + [b + d]_2 i, \\
	(a + bi)(c + di) & = [ac - bd]_2 + [ad + bc]_2 i,
\end{align*}
where $[\cdot]_2$ denotes the mod $2$ operation. Then the function $f_2(W_1,
W_2) = W_1 + i W_2$ is able to resolve all the ambiguities in
Fig.~\ref{fig:qpsk11}(b). Moreover, the function $f_2$ works well even
under other channel gains. In other words, the finite ring $\ZZ_2[i]$
seems to be a ``good match'' for QPSK constellation. This is not a
coincidence. As we will see later, every nested-lattice-based
constellation has such a good match.

\section{Algebraic Preliminaries}
\label{sec:preli}

In this section we recall some essential facts about principal ideal
domains, modules, and the Smith normal form, all of which will be useful
for our study of the algebraic properties of complex nested lattices.
All of this material is standard; see, e.g., \cite{M84,DF04,Brown93}.
We also introduce basic concepts and notation about lattices, mainly
based on \cite{Zamir-book, CS99}.

\subsection{Rings and Ideals}

We begin with some common definitions and notations for rings. All
rings in this paper will be commutative with identity $1 \neq 0$. Let
$R$ be a ring. We will let $R^*$ denote the nonzero elements of $R$,
i.e., $R^* = R \setminus \{ 0 \}$. An element $a$ is a \emph{divisor}
of an element $b$ in $R$, written $a \mid b$, if $b = ac$ for some
element $c \in R$. An element $u \in R$ is called a \emph{unit} of $R$
if $u \mid 1$. A non-unit element $p \in R$ is called a \emph{prime} of
$R$ if whenever $p \mid ab$ for some elements $a$ and $b$ in $R$, then
either $p\mid a$ or $p\mid b$. An element $a$ of $R^*$ is a called a
\emph{zero-divisor} if $ab = 0$ for some $b \in R^*$. If $R$ contains
no zero-divisors, then $R$ is an \emph{integral domain}.

An \emph{ideal} of $R$ is a nonempty subset $I$ of $R$ that is closed
under addition and inside-outside multiplication, i.e., for all $a,b \in
I$, $a+b \in I$ and for all $a \in I$ and all $r \in R$, $ar \in I$. If
$A$ is any nonempty subset of $R$, let $\langle A \rangle$ be the
smallest ideal of $R$ containing $A$, called the \emph{ideal generated
by $A$}. An ideal generated by a single element is called a
\emph{principal ideal}. A ring in which every ideal is principal is
called a \emph{principal ideal ring} (PIR).

Let $R$ be a ring and let $I$ be an ideal of $R$. Two elements $a$ and
$b$ are said to be \emph{congruent} modulo $I$ if $a-b \in I$.
Congruence modulo $I$ is an equivalence relation whose equivalence
classes are (additive) cosets $a + I$ of $I$ in $R$. The \emph{quotient
ring} of $R$ by $I$, denoted $R/I$, is the ring obtained by defining
addition and multiplication operations on the cosets of $I$ in $R$ in
the usual way, as
\[
 (a + I) + (b + I) = (a+b) + I \mbox{ and }
 (a + I) \times (b + I) = (ab) + I .
\]

\subsection{Principal Ideal Domains}

An integral domain in which every ideal is principal is called a
\emph{principal ideal domain} (PID). The integers $\ZZ$ form a PID. In
the context of complex lattices, typical examples of a PID include the
Gaussian integers $\ZZ[i]$ and the Eisenstein integers $\ZZ[\omega]$,
where $\omega = e^{2 \pi i / 3}$. Formally, Gaussian integers are the
set $\ZZ[i] \triangleq \{ a + bi : a, b \in \ZZ \}$, and Eisenstein
integers are the set $\ZZ[\omega] \triangleq \{ a + b\omega : a, b \in
\ZZ \}$.

The Gaussian integers $\ZZ[i]$ have four units ($\pm 1, \pm i$). A
Gaussian integer is called a {\em Gaussian prime} if it is a prime in
$\ZZ[i]$. A Gaussian integer $a+bi$ is a Gaussian prime if and only if
it satisfies exactly one of the following:
\begin{enumerate}
\item $|a| = |b| = 1$;
\item one of $|a|, |b|$ is zero and the other is a prime number in $\ZZ$
of the form $4j + 3$ (with $j$ a nonnegative integer);
\item both of $|a|, |b|$ are nonzero and $a^2 + b^2$ is a prime number
in $\ZZ$ of the form $4j + 1$.
\end{enumerate}
Note that these properties are symmetric with respect to $|a|$ and
$|b|$. Thus, if $a + bi$ is a Gaussian prime, so are $\{ \pm a \pm b i
\}$ and $\{ \pm b \pm a i \}$.

The Eisenstein integers $\ZZ[\omega] $ have six units ($\pm 1, \pm
\omega, \pm \omega^2$). An Eisenstein integer is called an {\em
Eisenstein prime} if it is a prime in $\ZZ[\omega]$. An Eisenstein
integer $a + b \omega$ is an Eisenstein prime if and only if it
satisfies exactly one of the following:
\begin{enumerate}
\item $a + b \omega$ is a product of a unit in $\ZZ[\omega]$ and a prime
number in $\ZZ$ of the form $3j + 2$;
\item $| a + b \omega |^2 = a^2 - ab + b^2$ is a prime number in $\ZZ$.
\end{enumerate}

Let $T$ be a PID and let $\pi \in T$. Then it is known
that the quotient $T / \langle \pi \rangle$ is a PIR
\cite{Brown93}.

\subsection{Modules}

Modules are to rings as vector spaces are to fields. Formally, let $R$
be a commutative ring with identity $1 \ne 0$. An $R$-module is a set
$M$ together with 1) a binary operation $+$ on $M$ under which $M$ is an
abelian group, and 2) an action of $R$ on $M$ which satisfies the same
axioms as those for vector spaces.

An $R$-submodule of $M$ is a subset of $M$ which itself forms an
$R$-module. Let $N$ be a submodule of $M$. The quotient group $M / N$
can be made into an $R$-module by defining an action of $R$ satisfying,
for all $r \in R$, and all $x+N \in M/N$, $r(x + N) = (rx) + N$. Hence,
$M / N$ is often referred to as a \emph{quotient $R$-module}.

Let $M$ and $N$ be $R$-modules. A map $\varphi: M \to N$ is called an
{\em $R$-module homomorphism} if the map $\varphi$ satisfies
\begin{enumerate}
\item $\varphi(x + y) = \varphi(x) + \varphi(y)$, for all $x, y \in M$ and
\item $\varphi(rx) = r\varphi(x)$, for all $r \in R, x \in M$.
\end{enumerate}
The {\em kernel of $\varphi$} is defined as $\ker \varphi \triangleq \{
m \in M : \varphi(m) = 0 \}$. Clearly, $\ker \varphi$ is a submodule of
$M$.

An $R$-module homomorphism $\varphi: M \to N$ is called an {\em
$R$-module isomorphism} if it is both injective and surjective. In this
case, the modules $M$ and $N$ are said to be {\em isomorphic}, denoted
by $M \cong N$. An $R$-module $M$ is called a \emph{free} module of
\emph{rank} $t$ if $M \cong R^t$ for some nonnegative integer $t$.

There are several isomorphism theorems for modules. The so-called
``first isomorphism theorem'' is useful for this paper.
\begin{thm}[First Isomorphism Theorem for Modules {\cite[p.~349]{DF04}}]
Let $M, N$ be $R$-modules and let $\varphi: M \to N$ be an $R$-module
homomorphism. Then $\ker \varphi$ is a submodule of $M$ and $M / \ker
\varphi \cong \varphi(M)$.
\end{thm}

\subsection{Modules over a PID}

Finitely-generated modules over PIDs play an important role in this
paper, and are defined as follows.

\begin{defi}[Finitely-Generated Modules]
Let $R$ be a commutative ring with identity $1 \ne 0$ and let $M$ be an
$R$-module. For any subset $A$ of $M$, let $\langle A \rangle$ be the
smallest submodule of $M$ containing $A$, called the \emph{submodule
generated by $A$}. If $M = \langle A \rangle$ for some finite subset
$A$, then $M$ is said to be {\em finitely generated}.
\end{defi}

A finite module (i.e., a module that contains finitely many elements) is
always finitely generated, but a finitely-generated module is not
necessarily finite. For example, the even integers $2\ZZ$ form a
$\ZZ$-module generated by $\{ 2 \}$.

The following structure theorem says that, if $T$ is a PID, then
a finitely-generated $T$-module is isomorphic to a finite direct
product of $T$-modules of the form $T$ or
$T/\langle \pi \rangle$.

\begin{thm}[Structure Theorem for Finitely-Generated Modules over a PID---Invariant Factor Form {\cite[p.~462]{DF04}}]\label{thm:fundamental}
Let $T$ be a PID and let $M$ be a finitely-generated
$T$-module. Then for some integer $t \ge 0$ and nonzero non-unit
elements $\pi_1, \ldots, \pi_k$ of $T$ satisfying the
divisibility relations $\pi_1 \mid \pi_2 \mid \cdots \mid \pi_k$,
\[
 M \cong T^t \times T/\langle \pi_1\rangle \times T/\langle \pi_2 \rangle \times
 \cdots \times T/\langle \pi_k \rangle.
\]
The elements $\pi_1, \ldots, \pi_k$, called the {\em invariant factors}
of $M$, are unique up to multiplication by units in $T$. The
integer $t$ is called the {\em free rank} of $M$.
\end{thm}

\subsection{Matrices over a PID}

Let $R^{m \times n}$ denote the set of all $m \times n$ matrices over
$R$. For any matrix $\bfA \in R^{m \times n}$, we denote by $a_{i,j}$
the entry at the $i$th row and $j$th column of $\bfA$. A matrix $\bfD
\in R^{m \times n}$ is called a \emph{diagonal matrix} if $d_{i,j} = 0$
whenever $i \ne j$. Note that a diagonal matrix need not be square. A
diagonal matrix $\bfD$ can be written as $\bfD = \diag(d_1, \ldots,
d_r)$, where $r = \min\{m, n\}$, and $d_i = d_{i,i}$ for $i = 1, \ldots,
r$.

A square matrix $\bfU \in R^{n \times n}$ is \emph{invertible} if $\bfU
\bfV = \bfV \bfU = \bfI_n$ for some $\bfV \in R^{n \times n}$, where
$\bfI_n$ denotes the ${n \times n}$ identity matrix. The set of
invertible matrices in $R^{n \times n}$, denoted as $\gl_n(R)$, forms a
group---the so-called \emph{general linear group}---under matrix
multiplication. Two matrices $\bfA, \bfB \in R^{m \times n}$ are said
to be \emph{equivalent} if there exist invertible matrices $\bfP \in
\gl_m(R)$ and $\bfQ \in \gl_n(R)$ such that $\bfB = \bfP \bfA \bfQ$. We
will write $\bfA \approx \bfB$ if $\bfA$ and $\bfB$ are equivalent.

\begin{defi}[Smith Normal Form]
Let $\bfA \in R^{m \times n}$ and let $r = \min\{ m, n\}$. A diagonal
matrix $\bfD = \diag(d_1, \ldots, d_r)$ is called a \emph{Smith normal
form} of $\bfA$ if $\bfD \approx \bfA$ and $d_1 \mid d_2 \mid \cdots
\mid d_r$ in $R$.
\end{defi}

Note that $d_1 \mid d_2 \mid \cdots \mid d_r$ in $R$ if and only if
$\langle d_1\rangle \supseteq \langle d_2\rangle \supseteq \cdots
\supseteq \langle d_r\rangle$. In particular, if $d_i$ is a unit in $R$,
then $d_1, \ldots, d_i$ are all units in $R$. Similarly, if $d_i = 0$,
then $d_i, \ldots, d_r$ are all $0$. Thus, if $\bfD = \diag(d_1,
\ldots, d_r)$ is a Smith normal form of $\bfA$, then the diagonal
entries $d_1, \ldots, d_r$ of $\bfD$ can be expressed as
\begin{equation*}
d_1, \ldots, d_r = \underbrace{u_1, \ldots, u_i}_i, \underbrace{d_{i+1}, \ldots,
d_{i+j}}_j ,\underbrace{0, \ldots, 0}_k
\end{equation*}
where $u_1, \ldots, u_i$ are units in $R$, $d_{i+1}, \ldots, d_{i+j}$
are nonzero, non-unit elements in $R$, and $i, j, k \ge 0$ with $i + j +
k = r$. The nonzero entries $\{u_1, \ldots, u_i, d_{i+1}, \ldots,
d_{i+j} \}$ are called a {\em sequence of invariant factors} of $\bfA$.

The Smith normal form theorem says that every matrix over a PID has a
Smith normal form whose sequence of invariant factors is unique up to
multiplication by units.
\begin{thm}[Smith Normal Form Theorem {\cite[p.~194]{Brown93}}]\label{thm:smith}
Let $T$ be a PID. Then any $\bfA \in T^{m \times n}$ has
a Smith normal form. Furthermore, if $\bfD_1 = \diag(d_1, \ldots, d_r)$
and $\bfD_2 = \diag(s_1, \ldots, s_r)$ are two Smith normal forms of
$\bfA$, then $\langle d_i\rangle = \langle s_i\rangle$ for all $i = 1,
\ldots, r$.
\end{thm}

\subsection{Lattices and Lattice Codes}

Recall that a real lattice $\Lambda \in \RR^n $ is a regular array of
points in $\RR^n$. Algebraically, a real lattice is defined as a
discrete $\ZZ$-submodule of $\RR^n$. A lattice $\Lambda \in \RR^n$ may
be specified by a set of $m$ basis (row) vectors $\bfg_1, \ldots, \bfg_m
\in \RR^n$, consisting of all $\ZZ$-linear combinations of the basis
vectors, i.e.,
\[
 \Lambda = \{ \bfr \bfG_{\Lambda} : \bfr \in \ZZ^m \},
\]
where $\bfG_{\Lambda} \triangleq \mat{\bfg_1^{T} | \cdots |
\bfg_m^{T}}^{T} \in \RR^{m \times n}$ is called a \emph{generator
matrix} for $\Lambda$. Note that $\bfG_{\Lambda}$ is not unique for a
given $\Lambda$. We call $m$ the {\em rank} of $\Lambda$, and $n$ the
{\em dimension} of $\Lambda$. Clearly, $m \le n$, because otherwise the
basis vectors cannot be linearly independent. When $m = n$, $\Lambda$
is called a {\em full-rank} real lattice.

Complex lattices are natural generalizations of real lattices. Let
$T$ be a discrete subring of $\CC$ forming a PID. Typical examples of
$T$ include the Gaussian integers $\ZZ[i]$ and the Eisenstein
integers $\ZZ[\omega]$. A \emph{$T$-lattice} $\Lambda$ in $\CC^n$ is
a discrete $T$-submodule of $\CC^n$, consisting of all $T$-linear
combinations of a set of basis vectors. Throughout this paper, we will
focus on full-rank $T$-lattices for simplicity, but all the results
can be easily extended to the case of non-full-rank $T$-lattices.

A few important notions are associated with a $T$-lattice. An
$n$-dimensional $T$-lattice $\Lambda$ partitions the space $\CC^n$
into \emph{congruent cells}. Such a partition is not unique. The most
important example is based on the \emph{nearest neighbor quantizer}
$\calQ_\Lambda^{\nn}$ that sends a point $\bfx \in \CC^n$ to a nearest
lattice point in Euclidean distance, {i.e.},
\begin{equation*}
 \calQ_\Lambda^{\nn}(\bfx) = \bflambda \in \Lambda, \quad
 \mbox{if }
 \forall \bflambda' \in \Lambda
 \left (\|\bfx - \bflambda\| \le \|\bfx - \bflambda'\| \right),
\end{equation*}
where ties are broken in a systematic manner. The \emph{Voronoi cell}
$\calV_{\Lambda}(\bflambda)$ associated with each $\bflambda \in
\Lambda$ is defined as the set of all points in $\CC^n$ that are closest
to $\bflambda$, i.e., $\calV_{\Lambda}(\bflambda) \triangleq \{\bfx \in
\CC^n: \calQ_\Lambda^{\nn}(\bfx)=\bflambda\}$. The cell
$\calV_{\Lambda}(\bfzero)$ associated with the origin is often referred
to as the \emph{Voronoi region} of $\Lambda$. Clearly, the Voronoi
cells $\{ \calV_{\Lambda}(\bflambda) \}$ have the following three
properties:
\begin{enumerate}
\item Each cell $\calV_{\Lambda}(\bflambda)$ is a shift of the cell
$\calV_{\Lambda}(\bfzero)$ by $\bflambda \in \Lambda$, i.e.,
$\calV_{\Lambda}(\bflambda) = \bflambda + \calV_{\Lambda}(\bfzero)$.
\item The cells do not intersect, i.e., $\calV_{\Lambda}(\bflambda) \cap
\calV_{\Lambda}(\bflambda') = \emptyset$ for all $\bflambda \ne
\bflambda'$.
\item The union of the cells covers the whole space, i.e.,
$\bigcup_{\bflambda \in \Lambda} \calV_{\Lambda}(\bflambda) = \CC^n$.
\end{enumerate}
In general, any collection of cells $\{ \calR_{\Lambda}(\bflambda) \}$
that satisfies the above three conditions is called a set of
\emph{fundamental cells}. The cell $\calR_{\Lambda}(\bfzero)$ associated
with the origin is called a \emph{fundamental region} and will also be
denoted simply by $\calR_{\Lambda}$. Note that every fundamental region
of a lattice $\Lambda$ has exactly the same volume, which is denoted by
$V(\Lambda)$.

A \emph{lattice quantizer} $\calQ_\Lambda: \CC^n \to \Lambda$
corresponding to $\calR_\Lambda$ sends every point $\bfx \in \CC^n$ to
the lattice point $\bflambda$ that is associated with the fundamental
cell $\calR_{\Lambda}(\bflambda)$ containing $\bfx$, i.e.,
\[
\calQ_\Lambda(\bfx) = \bflambda \in \Lambda, \
\mbox{if } \bfx \in \calR_{\Lambda}(\bflambda).
\]
Hence, any point $\bfx$ in $\CC^n$ can be uniquely expressed as the sum
of a lattice point and a point in the fundamental region
$\calR_{\Lambda}$, i.e., $\bfx = \calQ_\Lambda(\bfx) + \left( \bfx -
\calQ_\Lambda(\bfx) \right)$, where $\bfx - \calQ_\Lambda(\bfx)$ is a
point in $\calR_{\Lambda}$. This implies that, for all lattice points
$\bflambda \in \Lambda$ and all vectors $\bfz \in \CC^n$,
\begin{equation}\label{eq:quantizer_pro}
\calQ_\Lambda(\bflambda+\bfz) = \bflambda + \calQ_\Lambda(\bfz).
\end{equation}

The modulo-$\Lambda$ operation is defined, for a fixed $\calQ_\Lambda$,
as
\begin{equation*}
	\bfx \bmod \Lambda = \bfx - \calQ_\Lambda(\bfx).
\end{equation*}
Clearly, the modulo-$\Lambda$ operation always outputs a point in the
fundamental region $\calR_{\Lambda}$. The modulo-$\Lambda$ operation
has a geometrical interpretation:
\begin{equation*}
\bfx \bmod \Lambda = (\bfx + \Lambda) \cap \calR_{\Lambda},
\end{equation*}
where the \emph{lattice shift} $\bfx + \Lambda$ is defined as $\bfx +
\Lambda = \{ \bfx + \bflambda : \bflambda \in \Lambda \}$.

A \emph{$T$-sublattice} $\Lambda'$ of $\Lambda$ is a subset of
$\Lambda$ which is itself a $T$-lattice. Two lattices $\Lambda'$ and
$\Lambda$ are said to be \emph{nested} if $\Lambda'$ is a sublattice of
$\Lambda$, i.e., $\Lambda' \subseteq \Lambda$.

For each $\bflambda \in \Lambda$, the lattice shift $\bflambda +
\Lambda'$ is a {coset} of $\Lambda'$ in $\Lambda$, and the point
$\bflambda \bmod \Lambda'$ is called the \emph{coset leader} of
$\bflambda + \Lambda'$. Two cosets $\bflambda_1 + \Lambda'$ and
$\bflambda_2 + \Lambda'$ are either identical (when $\bflambda_1 -
\bflambda_2 \in \Lambda'$) or disjoint (when $\bflambda_1 - \bflambda_2
\notin \Lambda'$). Thus, the set of all distinct cosets of $\Lambda'$ in
$\Lambda$, denoted by $\Lambda / \Lambda'$, forms a partition of
$\Lambda$. Algebraically, $\Lambda / \Lambda'$ is a quotient
$T$-module, hereafter called a \emph{$T$-lattice quotient}.

A \emph{nested lattice code} $\calL(\Lambda,\Lambda')$ is defined as the
set of all coset leaders in $\Lambda / \Lambda'$, i.e.,
\[
	\calL(\Lambda,
\Lambda') = \Lambda \bmod \Lambda' = \{ \bflambda \bmod \Lambda' : \bflambda \in \Lambda \}.
\]
Geometrically, $\calL(\Lambda,\Lambda')$ is the intersection of the
lattice $\Lambda$ with the fundamental region $\calR_{\Lambda'}$, i.e.,
\[
	\calL(\Lambda,\Lambda') = \Lambda \cap \calR_{\Lambda'}.
\]
For this reason, the fundamental region $\calR_{\Lambda'}$ is often
interpreted as the \emph{shaping region}. Note that there is a bijection
between $\Lambda / \Lambda'$ and $\calL(\Lambda,\Lambda')$; in
particular,
\[
 |\Lambda / \Lambda'| = |\calL(\Lambda,\Lambda')| = V(\Lambda')/V(\Lambda).
\]

Finally, we mention that, for reasons of energy-efficiency, it is often
useful to consider a translated version of nested lattice codes. For
any fixed translation vector $\bfd \in \CC^n$, a \emph{translated nested
lattice code} $\calL(\Lambda, \Lambda', \bfd)$ is defined as
\[
	\calL(\Lambda, \Lambda', \bfd) = (\bfd + \Lambda) \bmod \Lambda' = (\bfd + \Lambda) \cap \calR_{\Lambda'}.
\]
\section{Problem Statement}
\label{sec:formulation}

This section gives a general definition of a \emph{linear}
physical-layer network coding (or compute-and-forward) scheme, and also
describes the assumptions on the system model made in this paper. We
focus on the problem faced by a receiver node of decoding one or more
linear combinations of simultaneously transmitted messages, as it is at
the heart of any system employing physical-layer network coding (see
\cite{NG11} for such a discussion). We conclude the section by briefly
describing some achievability results obtained by Nazer and Gastpar in
\cite{NG09}.

While linear network coding is traditionally defined over a finite field
\cite{LYC03,KM03}, our description considers a more general notion of
linear network coding over a finite commutative ring~$R$. In this
context, the message space, i.e., the set from where message packets are
drawn, is no longer a vector space, but an $R$-module \cite{DFZ05}. As
hinted at in Sec.~\ref{sec:motivating} and as will become clear in
Sec.~\ref{sec:lnc}, ring-linear network coding is required if we wish to
ensure compatibility with a \emph{general} lattice network coding
scheme.

\subsection{System Model}
\label{ssec:system-model}

\begin{figure}[t]
\centering\includegraphics[scale=0.5]{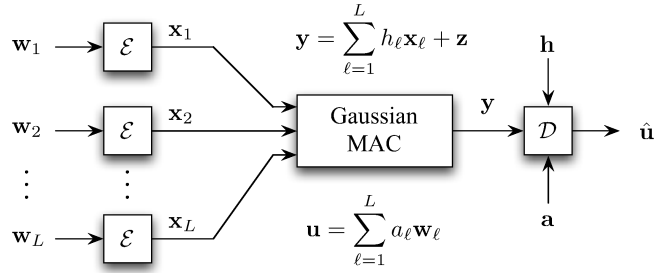}
\caption{Computing a linear function over a Gaussian
multiple-access channel.}
\label{fig:mac}
\end{figure}

Consider a multiple-access channel with $L$ transmitters and a single
receiver subject to block fading and additive white Gaussian noise, as
illustrated in Fig.~\ref{fig:mac}.

Channel inputs are denoted by $\bfx_1,\ldots,\bfx_L \in \CC^n$ and the
channel output is given by
\begin{equation*}
 \bfy = \sum_{\ell = 1}^L h_{\ell} \bfx_\ell + \bfz
\end{equation*}
where $h_1,\ldots,h_L \in \CC$ are channel gains (fading coefficients)
and $\bfz \sim \mathcal{CN}(\bfzero, N_0 \bfI_n)$ is a circularly-symmetric
jointly-Gaussian complex random vector. We assume that the channel gains
are perfectly known at the receiver but are \emph{unknown} at the
transmitters.

Transmitter $\ell$ is subject to a power constraint given by
\begin{equation*}
 \frac{1}{n}E\left[\| \bfx_\ell\| ^2\right] \leq P_\ell
\end{equation*}
where the expectation is taken with respect to a uniform distribution
over the corresponding message space. For simplicity (and without loss
of generality), we assume that the power constraint is symmetric, $P_1 =
\cdots = P_L \triangleq P$, and that any asymmetric power constraints
are incorporated by appropriately scaling the channel gains $h_\ell$.

For convenience, we define
\begin{equation*}
 \SNR \triangleq P/N_0.
\end{equation*}
Note that the received SNR corresponding to signal $\bfx_\ell$ is equal
to $|h_\ell|^2P/N_0$. Hence, the interpretation of $\SNR$ as the average
received SNR is only valid when $E[|h_\ell|^2] = 1$.

\subsection{Linear Physical-Layer Network Coding}

Let $R$ be a \emph{finite} commutative ring with identity $1 \ne 0$ and
let $T$ be some (usually infinite) commutative ring such that there
exists a surjective ring homomorphism $\sigma: T \to R$. Let the
\emph{ambient space} $W$ be a finite $R$-module. Note that $\sigma$
automatically makes $W$ into a $T$-module by defining $a \bfw =
\sigma(a)\bfw$, for all $a \in T$ and all $\bfw \in W$. As an example, we
may have $T = \ZZ$, $R = \ZZ/\langle 2 \rangle$, $W = \ZZ/\langle 2
\rangle$, and $\sigma(a) = a + \langle 2 \rangle$. In the following
setup, ``digital-layer'' network coding operates on $W$ over $R$, while
physical-layer network coding operates on $W$ over $T$, and the ring
homomorphism $\sigma$ guarantees the compatibility of such operations.

For each $\ell \in \{ 1,\ldots,L\}$, let the \emph{message space} of
transmitter~$\ell$ be an $R$-submodule $W_\ell \subseteq W$. A
$T$-linear PNC scheme with block length $n$ consists of $L$ encoders
\[
 \calE_\ell: W_\ell \to \CC^n
\]
each taking a message vector $\bfw_\ell \in W_\ell$ to a signal vector
$\bfx_\ell \in \CC^n$, and a decoder
\[
 \calD: \CC^n \to W
\]
that takes a received signal $\bfy \in \CC^n$ and attempts to compute one
(or more) $T$-linear combination(s) of the messages, such as
\[
 \bfu = \sum_{\ell = 1}^L a_{\ell} \bfw_\ell \in W
\]
whose coefficients $a_{\ell} \in T$ may or may not have been specified
\emph{a priori}. It is understood that any $T$-linear combinations
computed by the decoder are subsequently delivered to the digital layer
as $R$-linear combinations, such as
\[
 \bfu = \sum_{\ell = 1}^L a_{\ell} \bfw_\ell = \sum_{\ell = 1}^L \sigma(a_{\ell}) \bfw_\ell\in W
\]
obtained by the application of $\sigma$ on each coefficient.

The above generic description of the decoder may be specialized
depending on the problem at hand. Specifically, any further information
given to the decoder (such as side information about the channel gains)
will be denoted as additional arguments to $\calD$. Similarly, any
further information provided by the decoder will be denoted as
additional outputs of $\calD$. Note that, in this paper, we always
assume that the channel-gain vector $\bfh \triangleq (h_1, \ldots, h_L)
\in \CC^L$ is perfectly known at the receiver.

For simplicity of notation, let $\bfW \in W^L$ be a matrix corresponding
to the vertical stacking of $\bfw_1,\ldots,\bfw_L \in W$, taken as row
vectors. If the coefficient vector $\bfa = (a_1,\ldots,a_L) \in T^L$ for
the desired linear combination is specified \emph{a priori}, we will
write
\[
 \calD: \CC^n \times \CC^L \times T^L \to W, \quad \hat{\bfu} = \calD(\bfy|\bfh,\bfa).
\]
In this case, a decoding error is made if $\hat{\bfu} \neq \bfa \bfW$. The
corresponding probability of error is denoted by $P_e(\bfh,\bfa)$. This
decoder is illustrated in Fig.~\ref{fig:mac}.

If no coefficient vectors are given \emph{a priori}, but instead are
required to computed ``on-the-fly'' by the receiver, then we will write
\[
 \calD: \CC^n \times \CC^L \to W^m \times T^{L m}
\]
\[
 (\hat{\bfu}_1,\ldots,\hat{\bfu}_m,\bfa_1,\ldots,\bfa_m) = \calD(\bfy|\bfh)
\]
where $m$ denotes the number of linear combinations computed. In this
case, a decoding error is made if $\hat{\bfu}_i \neq \bfa_i \bfW$, for some
$i \in \{ 1,\ldots,m \}$.

Since a message is transmitted over $n$ (complex) channel uses, we
define the \emph{message rate} (spectral efficiency) for
transmitter~$\ell$ as $\Rmes{\;\!\!}_{,\ell} \triangleq
\frac{1}{n}\log_2|W_\ell|$, measured in bits per complex dimension.
Throughout the paper we assume that all encoders are identical, $\calE_1
= \ldots = \calE_\ell \triangleq \calE$, thus there is a single message
space~$W$ with message rate
\begin{equation*}
 \Rmes \triangleq \frac{1}{n}\log_2|W|.
\end{equation*}

As the following examples illustrate, a number of existing PNC schemes
can be described in this framework.

\begin{exam}
Let $L = 2$, $n=1$, $T = \ZZ$ and $R = W = \ZZ/\langle 2 \rangle$.
Consider the encoder
\[
 \calE(w) = \gamma\left(\emb{\sigma}(w) - \frac{1}{2}\right), \ w \in \ZZ/\langle 2 \rangle
\]
where $\gamma > 0$ is a scaling factor, and $\emb{\sigma}: \ZZ/\langle
2 \rangle \to \ZZ$ is defined as
\[
 \emb{\sigma}(w) = \begin{cases}
 1, & \text{when $w = 1 + \langle 2 \rangle$} \\
 0, & \text{when $w = 0 + \langle 2 \rangle$}.
\end{cases}
\]
Suppose $\bfh = [1 \ 1] \in \CC^2$. Let $\bfa = [1 \ 1] \in \ZZ^2$ be a
fixed coefficient vector. Then a decoder can be constructed as
\[
 \calD(y | \bfh,\bfa) = \begin{cases}
 1 + \langle 2 \rangle, & \text{if $|\Re\{y\}| < \gamma/2$} \\
 0 + \langle 2 \rangle, & \text{otherwise}.
 \end{cases}
\]
This is the simplest form of PNC \cite{ZLL06,PY06}, which may be
understood as XOR decoding under BPSK modulation, in the case of two
users with equal channel gains.
\end{exam}

\begin{exam}\label{exam:QAM}
 Let $L=2$, $n=1$, $T = \ZZ[i]$ and $R = W = \ZZ[i]/\langle m \rangle$, where $m$ is some
 positive integer. Consider the encoder
\[
 \calE(w) = \gamma\left(\emb{\sigma}(w) - d \right), \ w \in \ZZ[i]/\langle m \rangle
\]
where $d = \left(\frac{m-1}{2}\right)(1+i)$, $\gamma > 0$ is a scaling
factor, and $\emb{\sigma}: \ZZ[i] / \langle m \rangle \to \ZZ[i]$ is defined as
\[
 	\emb{\sigma}(a + bi + \langle m \rangle) = (a \bmod m) + (b \bmod m)i.
\]
First, suppose $\bfh = [1 \ 1] \in \CC^2$. Let $\bfa = [1 \ 1] \in
\ZZ[i]^2$ be the fixed coefficient vector. Then a natural (although
suboptimal) decoder is given by
\[
 \calD(y | \bfh,\bfa) = \left( \lfloor \Re\{y'\} \rceil
\bmod m \right) + 	\left( \lfloor \Im\{y'\} \rceil
\bmod m \right)i + \langle m \rangle,
\]
where $y' = {y}/{\gamma} + (a_1 + a_2)d$ and $\lfloor \cdot \rceil$
denotes the rounding operation. This scheme is known as the $m^2$-QAM
PNC scheme \cite{ZLL06}. Next, suppose $\bfh = [1 \ i] \in \CC^2$. Let
$\bfa = [1 \ i] \in \ZZ[i]^2$ be the fixed coefficient vector. Then the
above decoder generalizes the example discussed in
Sec.~\ref{sec:motivating}.
\end{exam}

\subsection{Achievable Rates}

We now mention some known achievable rates for the case of a single
given coefficient vector, under the assumptions of
Section~\ref{ssec:system-model}. These results were obtained by Nazer
and Gastpar \cite{NG09}.

\begin{thm}[\!\!\!{\cite{NG09}}]\label{thm:NG09}
For all~$\epsilon > 0$, all sufficiently large~$n$, and some
appropriately chosen prime integer $p$, there exists a $\ZZ[i]$-linear
PNC scheme with block length $n$ satisfying the following properties:
\begin{enumerate}
\item the message space is $W = \left(\ZZ[i]/\langle p \rangle
\right)^k$ for some $k$;
\item for any channel-gain vector $\bfh \in \CC^L$ and any non-zero
coefficient vector $\bfa \in \ZZ[i]^L$, the probability of decoding
error $P_e(\bfh, \bfa)$ is smaller than $\epsilon$ if $k$ is such that
the message rate $\Rmes$ is smaller than the computation rate
\[
 R_{\sf comp}(\bfh, \bfa) \triangleq \max_{\alpha \in \CC}
	\log_2\left( \frac{\SNR}{\| \alpha \bfh - \bfa \|^2 \SNR + |\alpha|^2} \right).
\]
\end{enumerate}
Moreover, the optimal value of $\alpha$ in the above expression is given
by
\begin{equation}\label{eq:opt_alpha}
	\alpha_\optt = \frac{\bfa \bfh^{\sf H} \SNR}{\| \bfh \|^2 \SNR + 1}
\end{equation}
which results in
\[
R_{\sf comp}(\bfh, \bfa) = \log_2\left( \frac{\SNR}{\bfa \bfM \bfa\hh}
 \right),
\]
where
\begin{equation}\label{eq:matrix-snr-gains}
 \bfM = \SNR \bfI_L - \frac{\SNR^2}{\SNR \| \bfh \|^2 + 1} \bfh^{\sf H}\bfh
\end{equation}
and $\bfI_L$ is the $L \times L$ identity matrix.
\end{thm}

\textit{Remark:} In the proof of the above result, $p$ has to grow
appropriately with $n$ such that $n/p \to 0$ as $n \to \infty$
\cite{NG09}.

Theorem~\ref{thm:NG09} is based on the existence of a ``good'' sequence
of nested lattices of increasing dimension. Criteria to design low
complexity, finite-dimensional PNC schemes are not immediately obvious
from these results. In the remainder of this paper, we will develop an
algebraic framework for studying linear PNC schemes, which facilitates
the construction and analysis of practical PNC schemes.

\section{Lattice Network Coding}
\label{sec:lnc}

\subsection{Linear Labelings}

Let $T$ be a discrete subring of $\CC$ forming a PID, and let
$\Lambda \subseteq \CC^n$ and $\Lambda' \subseteq \Lambda$ be two
full-rank $T$-lattices (called \emph{fine} and \emph{coarse},
respectively) so that the index $|\Lambda/\Lambda'|$ of $\Lambda'$ in
$\Lambda$ is finite. Recall that $\Lambda/\Lambda'$ is a {quotient
$T$-module}, i.e., it is a set closed under addition and
multiplication by elements of $T$. Specifically, addition of cosets
is defined as $(\bflambda_1 + \Lambda') + (\bflambda_2 + \Lambda')
\triangleq (\bflambda_1+\bflambda_2 + \Lambda')$, for all
$\bflambda_1,\bflambda_2 \in \Lambda$, multiplication by $r \in T$ is
defined as $r(\bflambda + \Lambda') \triangleq (r \bflambda +
\Lambda')$, for all $\bflambda \in \Lambda$, and multiplication
distributes over addition. An immediate consequence is that $\sum_{\ell
= 1}^L r_\ell(\bflambda_\ell + \Lambda') = (\sum_{\ell = 1}^L r_\ell
\bflambda_\ell) + \Lambda'$, i.e., a $T$-linear combination of cosets
is determined by the linear combination of their coset representatives.
This is the main property exploited in a lattice network coding (LNC)
scheme.

Conceptually, an LNC scheme is a $T$-linear PNC scheme based on a
finite lattice quotient $\Lambda/\Lambda'$, in which each transmitter
sends an information-embedding coset through a coset representative, and
each receiver recovers one or more $T$-linear combinations of the
transmitted coset representatives (which can potentially be forwarded to
other nodes according to the same scheme). Upon receiving enough such
combinations, the destination is able to decode all
information-embedding cosets from the transmitters.

To facilitate practical implementation, we will specify a map $\varphi:
\Lambda \to W$ from lattice points in $\Lambda$ to messages in the
message space $W$ for use in the above architecture. The map $\varphi$
must satisfy two conditions:
\begin{enumerate}
\item all points in the same coset are mapped to the same message, i.e.,
if for any two points $\bflambda_1, \bflambda_2 \in \Lambda$ with
$\bflambda_1 - \bflambda_2 \in \Lambda'$, $\varphi(\bflambda_1) =
\varphi(\bflambda_2)$;
\item the map $\varphi$ is $T$-linear, i.e., for all $r_1, r_2 \in
T$ and all $\bflambda_1, \bflambda_2 \in \Lambda$, we have
$\varphi\left( r_1 \bflambda_1 + r_2 \bflambda_2 \right) = r_1
\varphi(\bflambda_1) + r_2 \varphi(\bflambda_2)$.
\end{enumerate}
We refer to the map $\varphi$ as a \emph{linear labeling} of $\Lambda$.
As we shall see, it is this linear labeling that induces a natural
compatibility between the $\CC$-linear arithmetic of the multiple access
channel observed by the receiver and the $T$-linear arithmetic
desired in the message space.

The existence of the aforementioned linear labeling is guaranteed by the
following theorem, which provides a \emph{canonical decomposition} for
any finite $T$-lattice quotient $\Lambda/\Lambda'$.

\begin{thm}
\label{thm:structural}
Let $T$ be a PID and let $\Lambda$ and $\Lambda' \subseteq \Lambda$
be $T$-lattices such that $|\Lambda/\Lambda'|$ is finite. Then, for
some nonzero, non-unit elements $\pi_1, \pi_2, \ldots, \pi_k \in T$
satisfying the divisibility relations $\pi_1 \mid \pi_2 \mid \cdots \mid
\pi_k$, we have
\begin{equation}\label{eq:canonical}
	\Lambda / \Lambda' \cong
 T/\langle \pi_1 \rangle \times T/\langle \pi_2\rangle \times \cdots \times T/\langle \pi_k \rangle.
\end{equation}
Moreover, there exists a surjective $T$-module homomorphism
$\varphi:\Lambda \to T / \langle \pi_1 \rangle \times \cdots \times
T / \langle \pi_k \rangle$ whose kernel is $\Lambda'$.
\end{thm}
\begin{IEEEproof}
The first statement follows from Theorem~\ref{thm:fundamental} since
$\Lambda / \Lambda'$ is a finite $T$-module. The second statement
then follows from the First Isomorphism Theorem \cite{DF04}.
\end{IEEEproof}

\begin{figure}[t]
\centering\includegraphics[scale=0.25]{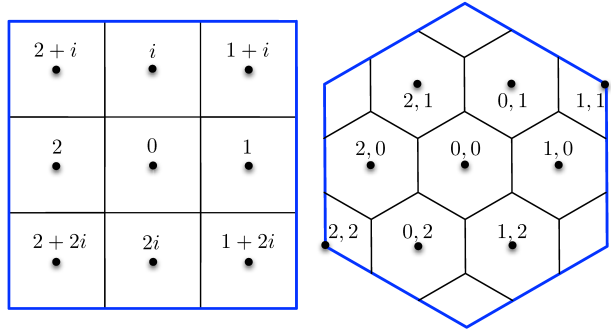}
\caption{Linear labelings for Examples~\ref{exam:G3} and \ref{exam:A2}.}
\label{fig:A2}
\end{figure}

Evidently, the map $\varphi$ is obtained as the composition of the
natural projection from $\Lambda$ to the quotient $\Lambda/\Lambda'$ with
the isomorphism of (\ref{eq:canonical}). According to
Theorem~\ref{thm:structural}, when the message space $W$ is taken as the
canonical decomposition in the right-hand side of (\ref{eq:canonical}),
i.e.,
\[
 W = T/\langle \pi_1 \rangle \times T/\langle \pi_2\rangle \times \cdots \times
 T/\langle \pi_k \rangle,
\]
the map $\varphi$ is indeed a linear labeling. The following examples
provide two concrete linear labelings, which are depicted in
Fig.~\ref{fig:A2}.

\begin{exam}\label{exam:G3}
Let $\Lambda = \ZZ[i]$ and $\Lambda' = 3 \ZZ[i]$. Let $T = \ZZ[i]$
and $W = \ZZ[i] / \langle 3 \rangle$. Consider the map $\varphi: \Lambda
\to W$ given by
\[
 \varphi(a + bi) = a + bi + \langle 3 \rangle.
\]
It is easy to check that the map $\varphi$ is $\ZZ[i]$-linear and its
kernel is $3 \ZZ[i]$.
\end{exam}

\begin{exam}\label{exam:A2}
Let $\Lambda$ be the (real) hexagonal lattice generated by $\bfg_1 = (1,
0)$ and $\bfg_2 = (1/2, \sqrt{3}/2)$. Let $\Lambda' = 3 \Lambda$. Let
$T = \ZZ$ and $W = \ZZ / \langle 3 \rangle \times \ZZ / \langle 3
\rangle$. Consider the map $\varphi: \Lambda \to W$ given by
\[
 \varphi(a \bfg_1 + b \bfg_2) = (a \bmod 3, b \bmod 3).
\]
It is easy to check that the map $\varphi$ is $\ZZ$-linear and its
kernel is $3\Lambda$.
\end{exam}

Linear labelings play a key role in LNC, as they directly map a
$T$-linear combination of transmitted lattice points to a
$T$-linear combination of transmitted messages, i.e., the latter can
be immediately extracted from the former.

It is also convenient to define an inverse operation, mapping a message
to a corresponding lattice point; this is done through an
\emph{embedding map} $\emb{\varphi}: W \to \Lambda$. This map must be an
injective function compatible with the linear labeling, so it must
satisfy
\[
 \varphi(\emb{\varphi}(\bfw)) = \bfw, \quad \text{for all $\bfw \in W$}.
\]

Equipped with a linear labeling $\varphi$ and and embedding
map~$\emb{\varphi}$, a high-level description of a generic LNC scheme
can be given as follows. Each encoder $\ell$ maps a message $\bfw_\ell
\in W$ to a lattice point $\bfx_\ell \in \Lambda$ labeled by
$\bfw_\ell$, i.e., $\bfx_\ell = \emb{\varphi}(\bfw_\ell)$. The decoder,
upon the reception of $\bfy$, and given a coefficient vector $\bfa =
(a_1,\ldots,a_L)$, attempts to compute the $T$-linear combination of
transmitted lattice points
\[
 \bflambda = \sum_{\ell = 1}^L a_\ell \bfx_\ell
\]
from which it would be able to extract the corresponding linear
combination of messages
\[
\bfu = \varphi(\bflambda) = \sum_{\ell = 1}^L a_\ell
 \varphi(\bfx_\ell)
 = \sum_{\ell = 1}^L a_\ell
\bfw_\ell.
\]

In more detail, the decoder proceeds in three steps. First, it scales
the received signal by a factor of $\alpha$, obtaining
\begin{equation}\label{eq:equivalent-channel}
 \alpha \bfy = \alpha \sum_{\ell = 1}^L h_\ell \bfx_\ell + \alpha \bfz = \bflambda + \bfn
\end{equation}
where
\begin{equation}\label{eq:effective-noise}
 \bfn = \sum_{\ell=1}^L (\alpha h_\ell - a_\ell) \bfx_\ell
+ \alpha \bfz
\end{equation}
is called the \emph{effective noise}. Note that we can view
(\ref{eq:equivalent-channel}) as an \emph{equivalent point-to-point
channel} under lattice coding: an effective message $\bfu$ is encoded as
a lattice point $\bflambda$, which is then additively corrupted by the
(signal-dependent and not necessarily Gaussian) effective noise $\bfn$.

Second, the decoder quantizes the scaled received signal with the fine
lattice to obtain
\begin{equation}\label{eq:quantizer}
 \hat{\bflambda} = \calQ_\Lambda(\alpha \bfy) = \calQ_\Lambda(\bflambda + \bfn) = \bflambda + \calQ_\Lambda(\bfn)
\end{equation}
where (\ref{eq:quantizer}) follows from the property
(\ref{eq:quantizer_pro}) of a lattice quantizer.

The last step is to apply the linear labeling, obtaining
\[
 \hat{\bfu} = \varphi(\hat{\bflambda}) = \varphi\left(\bflambda + \calQ_\Lambda(\bfn)\right) = \bfu + \varphi\left(\calQ_\Lambda(\bfn)\right).
\]
The decoder makes an error if and only if
$\varphi\left(\calQ_\Lambda(\bfn)\right) = \bfzero$ and therefore if and
only if $\calQ_{\Lambda}(\bfn) \in \Lambda'$. This is intuitive: if
$\calQ_{\Lambda}(\bfn) \in \Lambda'$, then the decoded lattice point
$\hat{\bflambda}$ is in the same coset as $\bflambda$ and is thus labeled
with~$\bfu$. On the other hand, if the decoded lattice point
$\hat{\bflambda}$ is labeled with~$\bfu$, then we must have
$\varphi(\calQ_{\Lambda}(\bfn)) = \bfzero$, which implies
$\calQ_{\Lambda}(\bfn) \in \Lambda'$, since the kernel of $\varphi$ is
$\Lambda'$.

\begin{figure}[t]
\centering\includegraphics[scale=0.3333]{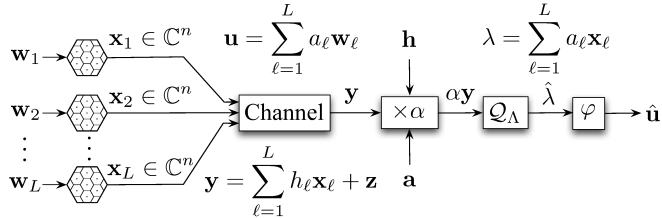}
\caption{Encoding and decoding architecture for LNC.}
\label{fig:mac2}
\end{figure}

To sum up, the above encoding-decoding architecture is depicted in
Fig.~\ref{fig:mac2}. The encoder $\calE: W \to \CC^n$ is given by
 \[
	\bfx_\ell = \calE(\bfw_\ell)
	= \emb{\varphi}(\bfw_\ell)
\]
and the decoder $\calD: \CC^n \times \CC^L \times T^L$ is given by
 \[
	\hat{\bfu} = \calD(\bfy | \bfh, \bfa) =
	\varphi(\calQ_{\Lambda}(\alpha \bfy))
\]
where $\alpha$ is a scaling factor chosen by the decoder based on $\bfh$
and $\bfa$, which will be discussed fully in the next section.
Intuitively, the purpose of $\alpha$ is to reduce the effective noise
$\bfn$, by trading off between \emph{self noise} (the first term in
(\ref{eq:effective-noise}) due to non-integer channel gains) and
Gaussian noise.

Clearly, the encoding-decoding complexity of an LNC scheme is not
essentially different from that for a point-to-point channel using the
same nested lattice code. Further, the error probability of the scheme
can be characterized by Proposition~\ref{prop:error}, as explained
before.

\begin{prop}\label{prop:error}
The message $\bfu = \sum_{\ell = 1}^L a_\ell \bfw_\ell$ is computed
incorrectly if and only if $\calQ_{\Lambda}(\bfn) \notin \Lambda'$.
That is, $\Pr[ \hat{\bfu} \ne \bfu] = \Pr[\calQ_{\Lambda}(\bfn) \notin
\Lambda' ]$.
\end{prop}

In practice, the nearest-neighbor quantizer $\calQ_{\Lambda}^{\nn}$ is
often preferred in the implementation of the decoder. This is to reduce
the error probability, as we will see in Sec.~\ref{sec:error_prob}.
Moreover, for reasons of energy-efficiency, a nested lattice code
$\calL(\Lambda, \Lambda')$ is usually preferred in the implementation of
the encoder. In this case, the encoder takes the messages in $W$ to
their \emph{minimum-energy} coset representatives, i.e., the embedding
map is chosen to satisfy
\[
	\emb{\varphi}(\bfw_\ell) = \emb{\varphi}(\bfw_\ell) \bmod \Lambda'
\]
where the shaping region $\calR_{\Lambda'}$ is chosen as the Voronoi
region.

Sometimes, a \emph{translated} nested lattice code can be used to
further reduce the energy consumption. Such techniques are well studied
in the area of Voronoi constellations (see, e.g., \cite{CS83,For89}).
Specifically, a translated version of a generic LNC scheme consists of
an encoder $\calE: W \times \CC^n \to \CC^n$
\begin{equation*}
\bfx_\ell = \calE(\bfw_\ell \mid \bfd_\ell)
 \triangleq \left( \bfd_\ell + \emb{\varphi}(\bfw_\ell) \right) \bmod \Lambda'
\end{equation*}
and a decoder $\calD: \CC^n \times \CC^L \times R^L \times (\CC^{n})^{L}
\to W$
\begin{equation*}
 \hat{\bfu} = \calD(\bfy\mid\bfh, \bfa, \{\bfd_\ell\})
 \triangleq \varphi\left( \calQ_{\Lambda}\left(\alpha \bfy - \sum_{\ell = 1}^L a_\ell \bfd_\ell \right) \right).
\end{equation*}
Note that Proposition~\ref{prop:error} holds unchanged in this case.

Finally, note that the message rate of an LNC scheme can be computed
geometrically as well as algebraically, as
\begin{align*}
\Rmes
&= \frac{1}{n} \log_2 \left(V(\Lambda') / V(\Lambda)\right) \\
&= \frac{1}{n} \sum_{i=1}^k \log_2 |T/\langle \pi_i \rangle|.
\end{align*}

\subsection{Construction of the Linear Labeling}

In this section, by applying the Smith normal form theorem, we provide
an explicit construction of the linear labeling $\varphi$ and an
embedding map $\emb{\varphi}$.

\begin{thm}\label{thm:construct}
Let $\Lambda / \Lambda'$ be a finite nested $T$-lattice quotient.
Then there exist generator matrices $\bfG_{\Lambda}$ and $\bfG_{\Lambda'}$
for $\Lambda$ and $\Lambda'$, respectively, satisfying
\begin{equation}\label{eq:generator-matrices-normal-form}
 \bfG_{\Lambda'} =
 \mat{{\rm diag}(\pi_1,\ldots,\pi_k) & \bfzero \\ \bfzero & \bfI_{n-k} }
 \bfG_{\Lambda}.
\end{equation}
In this case,
\begin{equation*}
 \Lambda / \Lambda' \cong T / \langle \pi_1 \rangle \times \cdots \times
 T / \langle \pi_k \rangle.
\end{equation*}
Moreover, the map
\[
 \varphi: \Lambda \to T / \langle \pi_1 \rangle \times \cdots \times
 T / \langle \pi_k \rangle
\]
given by
\begin{equation*}
 \varphi(\bfr \bfG_{\Lambda}) = (r_1 + \langle \pi_1 \rangle, \ldots, r_k +
 \langle \pi_k \rangle )
\end{equation*}
is a surjective $T$-module homomorphism with kernel $\Lambda'$.
\end{thm}
\begin{IEEEproof}
Let $\tilde{\bfG}_{\Lambda}$ and $\tilde{\bfG}_{\Lambda'}$ be any generator
matrices for $\Lambda$ and $\Lambda'$, respectively. Then
$\tilde{\bfG}_{\Lambda'} = \bfJ \tilde{\bfG}_{\Lambda}$, for some nonsingular
matrix $\bfJ \in {T^{n \times n}}$. Since $T$ is a PID, by
Theorem~\ref{thm:smith}, the matrix $\bfJ$ has a Smith normal form $\bfD
= \diag(d_1, \ldots, d_n)$. Since $\bfJ$ is nonsingular, the diagonal
entries $d_1, \ldots, d_n$ of $\bfD$ are all nonzero. Thus, $d_1,
\ldots, d_n$ can be expressed as
\[
	d_1, \ldots, d_n = u_1, \ldots, u_{n - k}, \pi_1, \ldots, \pi_k
\]
where $u_1, \ldots, u_{n - k}$ are units in $T$, $\pi_1, \ldots,
\pi_k$ are nonzero, non-unit elements in $T$. It follows that
\[
	\bfD \approx \tilde{\bfD} \triangleq
	 \mat{{\diag}(\pi_1,\ldots,\pi_k) & \bfzero \\ \bfzero & \bfI_{n-k} }.
\]
Therefore, $\bfJ \approx \tilde{\bfD}$ and there exist invertible
matrices $\bfP, \bfQ \in \gl_n(T)$ such that $\tilde{\bfD} = \bfP
\bfJ \bfQ$. We take
\begin{align*}
\bfG_\Lambda &= \bfQ^{-1} \tilde{\bfG}_{\Lambda} \\
\bfG_{\Lambda'} &= \bfP \tilde{\bfG}_{\Lambda'}
\end{align*}
as new generator matrices for $\Lambda$ and $\Lambda'$. Clearly, we
have $\bfG_{\Lambda'} = \tilde{\bfD} \bfG_{\Lambda}$. This proves the
first statement.

Since the second statement follows immediately from the third statement
and the First Isomorphism Theory, we need only to prove the third
statement here. That is, we must show that the map $\varphi$ is a
surjective $T$-homomorphism with kernel $\Lambda'$. Since it is easy
to check that the map $\varphi$ is surjective and $T$-linear, we will
show that the kernel of $\varphi$ is $\Lambda'$. Note that
\[
\varphi(\bfr\bfG_{\Lambda}) = \bfzero \iff
\forall i \in \{ 1, \ldots, k \}
 r_i \in \langle \pi_i \rangle.
\]
Note also that
\[
\Lambda' = \{ \bfr \bfG_{\Lambda} : r_i \in \langle \pi_i \rangle \},
\]
because $\bfG_{\Lambda'} = \tilde{\bfD} \bfG_{\Lambda}$. Hence, the kernel of
$\varphi$ is indeed $\Lambda'$.
\end{IEEEproof}

Theorem~\ref{thm:construct} constructs a linear labeling $\varphi:
\Lambda \to W$ explicitly. The key step is to find two generator
matrices $\bfG_{\Lambda}$ and $\bfG_{\Lambda'}$ satisfying the relation
(\ref{eq:generator-matrices-normal-form}). This can be achieved by
using the Smith normal form theorem. To construct an embedding map
$\emb{\varphi}$, one shall find a pre-image for each message $\bfw =
(r_1 + \langle \pi_1 \rangle, \ldots, r_k + \langle \pi_k \rangle)$.
Clearly, one natural choice of $\emb{\varphi}(\bfw)$ is given by
\[
\emb{\varphi}(\bfw) = (r_1, \ldots, r_k, \underbrace{0, \ldots, 0}_{n - k})\bfG_{\Lambda},
\]
which provides an explicit expression for $\emb{\varphi}(\bfw)$.

The use of the Smith normal form in coding theory is not new. In the
work of Forney \cite{For70, For89}, it was applied to study the
structure of convolutional codes as well as the linear labeling for real
lattices. The goal of the Smith normal form theorem is to reduce an
arbitrary matrix to a diagonal matrix, whose diagonal entries are the
invariant factors. In the context of complex $T$-lattices, such a
diagonal matrix reveals the nesting structure between the fine lattice
and the coarse lattice, leading to a transparent linear labeling.

\subsection{End-to-End Perspective}

In this section, we study the use of LNC in a non-coherent network model
(where destinations have no knowledge of the operations of relay nodes)
rather than the coherent network model described in \cite{NG09}. To
provide a context for our study, we consider a Gaussian relay network in
which a generic LNC scheme is used in conjunction with a scheduling
algorithm. The scheduling algorithm indicates, at each time slot, which
nodes are transmitters and which nodes are receivers. As a transmitter,
a node first computes a random linear combination of the packets in its
buffer and then maps this combination to a transmitted signal. As a
receiver, a node first decodes the received signal into one or more
linear combinations of the transmitted packets and then performs (some
form of) Gaussian elimination in order to discard redundant (linearly
dependent) packets in the buffer.

Initially, only the source nodes have nonempty buffers containing the
message packets. When the communication ends, each destination node
will have collected sufficiently many linear combinations of the
message packets. This induces an end-to-end linear network-coding
channel in which the message space $W$ is, in general, a $T$-module
$T / \langle \pi_1 \rangle \times \cdots \times T / \langle \pi_k
\rangle$. Since modules over PIDs share much in common with vector
spaces over finite fields, it would be natural to expect that many
useful techniques for non-coherent network coding can be adapted here.

We use the technique of headers as an illustrating example in this
section. For convenience, we rewrite the message space as
\[
 	W = T / \langle \pi_k \rangle \times \cdots \times T / \langle \pi_1 \rangle.
\]
Similar to the vector-space case, we use the first $m$ components to
store headers, and the last $k - m$ components to store payloads, where
$m$ is the number of message packets. Specifically, the header for the
$i$th message packet is a length-$m$ tuple with $1 + \langle \pi_{k -i +
1} \rangle$ at position $i$ and $0 + \langle \pi_{k - j + 1} \rangle$ at
other positions (where $1 \le j \le m$ and $j \ne i$).

\begin{exam}\label{exam:message-matrix}
Let the message space $W = \ZZ / \langle 12 \rangle \times \ZZ / \langle
6 \rangle \times \ZZ / \langle 2 \rangle \times \ZZ / \langle 2
\rangle$. Suppose there are $2$ original messages in the system. Then
the matrix $\bfW$ of the source messages is of the form
\[
		\bfW = \mat{1 + \langle 12 \rangle & 0+\langle 6 \rangle &
		a + \langle 2 \rangle & b + \langle 2 \rangle \\ 0+\langle 12 \rangle &
		1 + \langle 6 \rangle & c + \langle 2 \rangle & d + \langle 2 \rangle },
	\]
where $a, b, c, d \in \ZZ$.
\end{exam}

Recall that, when the message space is a vector space, Gauss-Jordan
elimination is used to recover the payloads at the destinations. As one
may expect, for a more general message space, some modification of
Gauss-Jordan elimination is needed. It turns out that the key step in
the modification is to transform a $2 \times 1$ matrix to a row echelon
form: given $a, b \in T$, return $s, t, u, v, g \in T$ such that
\[
	\mat{ s & t \\ u & v} \mat{a \\ b} = \mat{g \\ 0}
\]
where the determinant, $sv - tu$, is a unit from $T$.
\begin{exam}\label{exam:header-decoding}
Suppose that the matrix $\bfW$ of the message packets is given in
Example~\ref{exam:message-matrix}. Suppose that a destination has
received two linear combinations, $2 \bfw_1 + 3 \bfw_2$ and $3\bfw_1 + 2
\bfw_2$. Then the matrix $\bfY$ of the received packets at the
destination is $\bfY = \mat{2 & 3 \\ 3 & 2} \bfW$, which is in the form
of
\[
 \bfY = \mat{2 + \langle 12 \rangle & 3 + \langle 6 \rangle & c + \langle 2 \rangle &
		d + \langle 2 \rangle \\ 3 + \langle 12 \rangle & 2 + \langle 6 \rangle &
		 a + \langle 2 \rangle & b + \langle 2 \rangle}.
\]
To recover the payloads, we reduce the first column of $\bfY$ to a row
echelon form. Since
\[
		\mat{2 & -1 \\ -3 & 2} \mat{2 \\ 3} = \mat{1 \\ 0}
\]
over $\ZZ$ and the determinant, $2\times 2 - (-1)\times(-3) = 1$, is a
unit in $\ZZ$, we multiply the matrix $\mat{2 & -1 \\ -3 & 2}$ with
$\bfY$, obtaining
\begin{align*}
		\bfY_1 &= \mat{2 & -1 \\ -3 & 2} \bfY \\
		&= \mat{1 + \langle 12 \rangle & 4 + \langle 6 \rangle
		 & a + \langle 2 \rangle & b + \langle 2 \rangle \\ 0 + \langle 12 \rangle &
		 1 + \langle 6 \rangle & c + \langle 2 \rangle & d + \langle 2 \rangle}.
\end{align*}
In this way, we transform the matrix $\bfY$ to a row echelon form.
Next, we transform the matrix $\bfY$ to a reduced row echelon form,
which can be done by subtracting $4$ times the second row from the first
row, i.e.,
\[
		\bfY_2 = \mat{1 & -4 \\ 0 & 1} \bfY_1.
\]
Now it is easy to check that $\bfY_2 = \bfW$. In other words, the
payloads are recovered correctly.
\end{exam}

Although Example~\ref{exam:header-decoding} only illustrates the
decoding procedure for the case of $m=2$, it can be extended to the case
of $m > 2$ through a simple mathematical induction.

Finally, we would like to point out that the design of headers in
Example~\ref{exam:message-matrix} is suboptimal, and a better design can
be made by using matrix canonical forms. The development of this idea
is beyond the scope of this paper and will instead be discussed in a
separate paper \cite{FNKS13-submitted}.

\section{Performance Analysis for Lattice Network Coding}\label{sec:error_prob}

In this section, we turn from algebra to geometry, presenting an
error-probability analysis as well as its implications.

\subsection{Error Probability for LNC}

Recall that, according to Proposition~\ref{prop:error}, the error
probability of decoding a linear function $\bfu$ is $\Pr[\hat{\bfu} \ne
\bfu] = \Pr[\mathcal{Q}_{\Lambda}(\bfn) \notin~\Lambda']$, where $\bfn$ is
the effective noise given by (\ref{eq:effective-noise}). Note that the
effective noise $\bfn$ is not necessarily Gaussian, making the analysis
nontrivial. To alleviate this difficulty, we focus on a special case in
which the shaping region $\calR_{\Lambda'}$ is a (rotated) hypercube in
$\CC^n$, {i.e.},
\begin{equation}\label{eq:hyper}
\calR_{\Lambda'} = \gamma \bfU \calH_n
\end{equation}
where $\gamma > 0$ is a scalar factor, $\bfU$ is any $n \times n$ unitary
matrix, and $\calH_n$ is a unit hypercube in $\CC^n$ defined by $\calH_n
= \left( [-1/2, 1/2) + i[-1/2, 1/2) \right)^n$. This case corresponds
to the so-called {\em hypercube shaping} in \cite{SFS09}. The
assumption of hypercube shaping not only simplifies the analysis of
error probability, but also has some practical advantages, for example,
the complexity of the shaping operation is generally low. However, as we
will see later, there is no shaping gain under hypercube shaping. This
is expected, since similar results hold for the use of lattice codes in
point-to-point channels \cite{For89, SFS09}.

In the sequel, we will provide an approximate upper bound for the error
probability for LNC schemes admitting hypercube shaping. This upper
bound is closely related to certain geometrical parameters of a lattice
quotient as defined below.

Let us define the \emph{minimum (inter-coset) distance} of a lattice
quotient $\Lambda / \Lambda'$ as
\begin{align*}
d(\Lambda / \Lambda')
&\triangleq \min_{\bflambda_1,\bflambda_2 \in \Lambda: \bflambda_1 - \bflambda_2 \not\in \Lambda'} ||\bflambda_1 - \bflambda_2|| \\
&= \min_{\bflambda \in \Lambda \setminus \Lambda'} ||\bflambda||
\end{align*}
where $\Lambda \setminus \Lambda'$ denotes the set difference $\{
\bflambda \in \Lambda : \bflambda \notin \Lambda' \}$. Note that
$d(\Lambda / \Lambda')$ corresponds to the length of the shortest
vectors in $\Lambda \setminus \Lambda'$. Let $K(\Lambda / \Lambda')$
denote the number of these shortest vectors.

We have the following union bound estimate on the error probability.

\begin{thm}[Probability of Decoding Error]\label{thm:criterion}
Suppose that the shaping region $\calR_{\Lambda'}$ is a (rotated)
hypercube and that all the transmitted vectors are independent and
uniformly distributed over $\calR_{\Lambda'}$. Suppose that
$\calQ_\Lambda(\cdot)$ is a nearest-neighbor quantizer. Then a union
bound estimate on the error probability in decoding a specified linear
combination is
\begin{align}
	&\ P_e(\bfh, \bfa) \nonumber \\
	&\lessapprox \min_{\alpha \in \CC}K(\Lambda / \Lambda') \exp\left(- \frac{d^2(\Lambda / \Lambda')}{4N_0 (|\alpha|^2 + \SNR \|\alpha \bfh - \bfa\|^2)}\right). \label{eq:error-prob}
\end{align}
Moreover, the optimal value of $\alpha$, i.e., the value of $\alpha$
that minimizes the right-hand side of (\ref{eq:error-prob}), is given by
(\ref{eq:opt_alpha}), which results in
\begin{equation}\label{eq:union-bound-estimate}
P_e(\bfh, \bfa) \lessapprox K(\Lambda / \Lambda') \exp\left(- \frac{d^2(\Lambda / \Lambda')}{4N_0
	\bfa \bfM \bfa\hh}\right)
\end{equation}
where the matrix $\bfM$ is given by (\ref{eq:matrix-snr-gains}).
\end{thm}
The proof is given in Appendix~\ref{app:error}. Note that the proof
assumes the use of random dithering (translation by a random vector
chosen uniformly at random from the shaping region) at the encoders, so
that the transmitted vectors are uniformly distributed over the shaping
region.

Theorem~\ref{thm:criterion} implies that the lattice quotient $\Lambda /
\Lambda'$ should be designed such that $K(\Lambda / \Lambda')$ is
minimized and $d(\Lambda / \Lambda')$ is maximized (under a given
message rate $\Rmes$ and $\SNR$), which will be discussed fully in
Sec.~\ref{sec:design}. Further, if the receiver has the freedom to
choose the coefficient vector $\bfa$, it needs to minimize the term
$\bfa \bfM \bfa\hh$, which, as observed in \cite{ZNGE09}, is a shortest
vector problem. Theorem~\ref{thm:criterion} can be extended to other
shaping methods. A particular example is provided in \cite{QJ12}.

\subsection{Nominal Coding Gain}

Similarly to the point-to-point case, we define the \emph{nominal coding
gain} of $\Lambda / \Lambda'$ as
\[
	\gamma_c(\Lambda / \Lambda') \triangleq \frac{d^2(\Lambda / \Lambda')}{V(\Lambda)^{1/n}}.
\]
Note that the nominal coding gain is invariant to scaling. For an LNC
scheme with hypercube shaping, we have $V(\Lambda') = \gamma^{2n}$ and
$P = \gamma^2 /6$ where $\gamma > 0$ is the scalar factor in
(\ref{eq:hyper}). Thus, $V(\Lambda')^{1/n} = 6P$. Note also that
$V(\Lambda)^{1/n} = 2^{-\Rmes} V(\Lambda')^{1/n}$. It follows that the
union bound estimate in (\ref{eq:union-bound-estimate}) can be expressed
as
\[
 P_e(\bfh, \bfa) \lessapprox K(\Lambda / \Lambda') \exp\left(- \frac{3}{2} \gamma_c(\Lambda / \Lambda') 2^{-\Rmes} \frac{\SNR}{\bfa \bfM \bfa^{\sf H}}\right).
\]
Thus, for a given spectral efficiency $\Rmes$, the performance of such
an LNC scheme can be characterized by the parameters $K(\Lambda /
\Lambda')$ and $\gamma_c(\Lambda / \Lambda')$.

Note that the nominal coding gain of a baseline lattice quotient
$\ZZ[i]^n/\pi\ZZ[i]^n$ is equal to $1$ for all $\pi \in \ZZ[i]^*$.
Thus, $\gamma_c(\Lambda / \Lambda')$ provides a first-order estimate of
the performance improvement of an LNC scheme over a baseline LNC scheme.
For this reason, $\gamma_c(\Lambda / \Lambda')$ will be used as a figure
of merit of LNC schemes in the rest of this paper; yet the effect of
$K(\Lambda / \Lambda')$ cannot be ignored in a more detailed assessment
of LNC schemes.

\section{Design of Nested Lattices}\label{sec:design}

In this section, we adapt several known lattice constructions to produce
pairs of nested lattices with simple message space and high coding gain.

\subsection{Constructions of Nested Lattices}\label{sec:new_construct}

Known methods for designing lattices include Construction~A and
Construction~D as well as their complex versions (see, e.g.,
\cite{CS99}). Here, we adapt these methods to construct pairs of nested
lattices. In all of our examples, the Voronoi region of the coarse
lattice is chosen as its fundamental region.

\subsubsection{Nested Lattices via Construction~A}
Let $p > 0$ be a prime number in $\ZZ$. Let $\calC$ be a linear code of
length $n$ over $\ZZ/ \langle p \rangle$. Without loss of generality, we
may assume the linear code $\calC$ is systematic. Define a ``real
Construction~A lattice'' \cite{CS99} as
\[
	\realL \triangleq
 \{ \bflambda \in \ZZ^n : \sigma(\bflambda)\in \calC \},
\]
where $\sigma: \ZZ^n \to (\ZZ/\langle p \rangle)^n$ is the natural
projection map. (Here, the subscript $r$ stands for ``real.'') Define
\[
	\realLc \triangleq \{ p \bfr : \bfr \in
	\ZZ^n \}.
\]
It is easy to see that $\realLc$ is a sublattice of $\realL$. Hence, we
obtain a pair of nested $\ZZ$-lattices $\realL \supseteq \realLc$ from
the linear code $\calC$.

Now we ``lift'' this pair of nested $\ZZ$-lattices to a pair of nested
$\ZZ[i]$-lattices. Let $\complexL = \realL + i \realL$, i.e.,
\[
\complexL = \{ \bflambda \in \ZZ[i]^n :
\Re\{\bflambda\}, \Im\{\bflambda\} \in \realL \}.
\]
Similarly, let $\complexLc = \realLc + i \realLc$. In this way, we
obtain a pair of nested $\ZZ[i]$-lattices $\complexL \supseteq
\complexLc$. A variant of this construction was used by Nazer and
Gastpar in \cite{NG09}.

To study the message space induced by $\complexL / \complexLc$, we
specify two generator matrices satisfying the relation
(\ref{eq:generator-matrices-normal-form}). On the one hand, we note
that the lattice $\realL$ has a generator matrix $\bfG_{\realL}$ given by
\[
	\bfG_{\realL} = \left[ \begin{array}{cc} \bfI_k &
	 \bfB_{k \times (n - k)} \\ \bfzero_{(n - k) \times k} & p \bfI_{n-k} \end{array} \right],
\]
where $\sigma([\bfI \ \bfB])$ is a generator matrix for $\calC$. The
lifted lattice $\complexL$ has a generator matrix $\bfG_{\complexL}$ that
is identical to $\bfG_{\realL}$, but over $\ZZ[i]$. On the other hand,
we note that the lattice $\complexLc$ has a generator matrix
$\bfG_{\complexLc}$ given by
\[
	\bfG_{\complexLc}= \left[ \begin{array}{cc} p \bfI_k &
	p \bfB_{k \times (n - k)} \\ \bfzero_{(n - k) \times k} &
	p \bfI_{n-k} \end{array} \right].
\]
These two generator matrices $\bfG_{\complexL}$ and $\bfG_{\complexLc}$
satisfy
\[
 \bfG_{\complexLc} =
 \mat{p \bfI_k & \bfzero \\ \bfzero & \bfI_{n-k} }
 \bfG_{\complexL}.
\]
It follows from Theorem~\ref{thm:construct} that $\complexL / \complexLc
\cong (\ZZ[i]/\langle p \rangle)^k$. That is, the message space under
this construction is $W = (\ZZ[i]/\langle p \rangle)^k$. In particular,
the message rate $\Rmes = \frac{k}{n} \log_2(p^2)$, since
$\ZZ[i]/\langle p \rangle$ contains $p^2$ elements.

Note that the message space $W$ can be viewed as a free $\ZZ[i]/\langle
p \rangle$-module of rank $k$. In particular, $W$ is a vector space if
and only if the prime number $p$ is a Gaussian prime, which is
equivalent to saying that $p$ is of the form $4j + 3$.

To study the nominal coding gain $\gamma_c(\complexL / \complexLc)$ as
well as $K(\complexL / \complexLc)$, we relate them to certain
parameters of the linear code $\calC$. To each codeword $\bfc = (c_1 +
\langle p \rangle, \ldots, c_n + \langle p \rangle) \in \calC$, there
corresponds a coset $(c_1, \ldots, c_n) + p \ZZ^n$ whose minimum-norm
coset leader, denoted by $\best{\bfc}$, is given by
\begin{equation*}
\best{\bfc} = (c_1 - \lfloor c_1/p \rceil \times p, \ldots, c_n - \lfloor c_n/p \rceil \times p),
\end{equation*}
where $\lfloor x \rceil$ is a rounding operation. The Euclidean weight
$w_E{(\bfc)}$ of $\bfc$ {can then be} defined as the squared Euclidean
norm of $\best{\bfc}$, that is, $w_E{(\bfc)} = \| \best{\bfc} \|^2$. Thus,
for example, when $\bfc = (1 + \langle 5 \rangle, 3 + \langle 5
\rangle)$, $\best{\bfc} = (1, -2)$. Clearly, the Euclidean weight of
$\bfc$ is equivalent to the $2$-norm of $\bfc$ defined in \cite{RS87}.
Let $w_E^{\min}(\calC)$ be the minimum Euclidean weight of nonzero
codewords in $\calC$, i.e.,
\[
w_E^{\min}(\calC) = \min\{ w_E{(\bfc)} : \bfc \ne \bfzero, \ \bfc \in \calC \}.
\]
Let $A(w_E^{\min})$ be the number of codewords in $\calC$ with minimum
Euclidean weight $w_E^{\min}(\calC)$. Then we have the following
result.
\begin{prop}\label{prop:construction-a}
Let $\calC$ be a linear code over $\ZZ/\langle p \rangle$ and let
$\complexL \supseteq \complexLc$ be a pair of nested lattices
constructed from $\calC$. Then
\[
\gamma_c(\complexL / \complexLc) = \frac{w_E^{\min}(\calC)}{p^{2(1-k/n)}}
\]
and
\[
K(\complexL / \complexLc) = \begin{cases}
 2A\left(w_E^{\min}(\calC)\right)2^{w_E^{\min}(\calC)}, & \text{when $p = 2$}, \\
 2A\left(w_E^{\min}(\calC)\right), & \text{when $p > 2$}.
 \end{cases}
\]
\end{prop}
The proof is in Appendix~\ref{app:construction-a}.

Proposition~\ref{prop:construction-a} suggests that optimizing the
nominal coding gain $\gamma_c(\complexL / \complexLc)$ amounts to
maximizing the minimum Euclidean weight $w_E^{\min}(\calC)$ of $\calC$,
and that optimizing $K(\complexL / \complexLc)$ amounts to minimizing
$A(w_E^{\min})$.

\subsubsection{Nested Lattices via Complex Construction~A}
Let $\pi$ be a prime in $T$. Let $\calC$ be a linear code of length
$n$ over $T / \langle {\pi} \rangle$. Without loss of generality, we
may assume the linear code $\calC$ is systematic. Define a ``complex
Construction~A lattice'' \cite{CS99} as
\[
 \complexL \triangleq
 \{ \bflambda \in T^n : \sigma(\bflambda) \in \calC \},
\]
where $\sigma: T^n \to (T / \langle \pi \rangle)^n$ is the natural
projection map. Define
\[
 \complexLc \triangleq \{ \pi \bfr : \bfr \in T^n \}.
\]
It is easy to see $\complexLc$ is a sublattice of $\complexL$. Hence, we
obtain a pair of nested lattices $\complexL \supseteq \complexLc$ from
the linear code $\calC$.

To study the message space induced by $\complexL / \complexLc$, we
specify two generator matrices satisfying the relation
(\ref{eq:generator-matrices-normal-form}). It is well-known that
$\complexL$ has a generator matrix $\bfG_{\complexL}$ given by
\[
	\bfG_{\complexL} = \left[
 \begin{array}{cc} \bfI_k & \bfB_{k \times (n - k)} \\
 \bfzero_{(n - k) \times k} & \pi \bfI_{n-k} \end{array}
 \right],
\]
and that $\complexLc$ has a generator matrix $\bfG_{\complexLc}$ given by
\[
	\bfG_{\complexLc}= \left[
 \begin{array}{cc} \pi \bfI_k & \pi \bfB_{k \times (n - k)} \\
 \bfzero_{(n - k) \times k} & \pi \bfI_{n-k} \end{array}
 \right].
\]
These two generator matrices satisfy
\[
 \bfG_{\complexLc} =
 \mat{\pi \bfI_k & \bfzero \\ \bfzero & \bfI_{n-k} }
 \bfG_{\complexL}.
\]
Hence, we have $\complexL / \complexLc \cong (T / \langle \pi
\rangle)^k$. That is, the message space under this construction is $W =
(T / \langle \pi \rangle)^k$. Since $\pi$ is a prime in $T$,
$T / \langle \pi \rangle$ is a finite field and $W$ is a vector space
of dimension $k$. Thus, this construction is preferable to the previous
construction, if the message space is required to be a vector space.
For instance, if $T = \ZZ[\omega]$ and $\pi = 2$, then the message
space $W$ is a vector space over $\FF_4$. This never happens under the
previous construction, since $2$ is not a prime in $\ZZ[i]$.

To study the nominal coding gain $\gamma_c(\complexL / \complexLc )$ as
well as $K(\complexL / \complexLc )$, we again relate them to the
parameters of the linear code $\calC$ with a particular focus on $T =
\ZZ[i]$ (due to hypercube shaping). The definition of the minimum
Euclidean weight $w_E^{\min}(\calC)$ is the same as the previous
definition, except for the fact that the minimum-norm coset leader
$\best{\bfc}$ is given by
\begin{equation*}
\best{\bfc} = (c_1 - \lfloor c_1/\pi \rceil \times \pi, \ldots, c_n - \lfloor c_n/\pi \rceil \times \pi),
\end{equation*}
where the rounding operation $\lfloor x \rceil$ sends $x \in \CC$ to the
closest Gaussian integer in the Euclidean distance.
\begin{prop}\label{prop:construction-aa}
Let $\calC$ be a linear code over $\ZZ[i]/\langle \pi \rangle$ and let
$\complexL \supseteq \complexLc$ be a pair of nested lattices
constructed from $\calC$. Then
\[
	\gamma_c(\complexL / \complexLc ) = \frac{w_E^{\min}(\calC)}{|\pi|^{2(1-k/n)}}
\]
and
\[
 K(\complexL / \complexLc ) = \begin{cases}
 A\left(w_E^{\min}(\calC)\right)4^{w_E^{\min}(\calC)}, & \text{when $|\pi|^2 = 2$}, \\
 A\left(w_E^{\min}(\calC)\right), & \text{otherwise}.
 \end{cases}
\]
\end{prop}
The proof is in Appendix~\ref{app:construction-aa}.

\subsubsection{Nested Lattices via Construction~D}

Let $p > 0$ be a prime in $\ZZ$. Let $\calC_1 \subseteq \cdots \subseteq
\calC_s$ be nested linear codes of length $n$ over $\ZZ/\langle p
\rangle$, where $\calC_i$ has parameters $[n, k_i]$ for $i = 1, \ldots,
s$. As shown in \cite{CS99}, there exists a basis $\{ \bfg_1, \ldots,
\bfg_n \}$ for the vector space $(\ZZ/\langle p \rangle)^n$ such that
\begin{enumerate}
 \item $\bfg_1, \ldots,
\bfg_{k_i}$ span $\calC_i$ for $i = 1, \ldots, s$; and
 \item if $\bfG$
denotes the matrix with rows $\bfg_1, \ldots, \bfg_n$, some permutation of
the rows of $\bfG$ gives an upper triangular matrix with diagonal
elements equal to $1 + \langle p \rangle$.
\end{enumerate}
(In fact, $\bfG$ can be constructed by applying Gaussian elimination to
the generator matrices of the nested linear codes iteratively.)

Using the nested linear codes $\{ \calC_i, 1 \le i \le s \}$, we define
a ``real Construction~D lattice'' \cite{CS99} as
\begin{align}
	\realL
	\triangleq&
	 \left\{ \sum_{i = 1}^s
	\sum_{j = 1}^{k_i} p^{i-1} \beta_{ij} \emb{\sigma}(\bfg_j) :
	\beta_{ij} \in \{0, \ldots, p-1 \} \right\}\nonumber \\
 &+ p^s \ZZ^n \label{eq:construction-D-real-lattice}
\end{align}
where $\emb{\sigma}$ is the natural embedding map from $(\ZZ/\langle p
\rangle)^n$ to $\{0, \ldots, p-1 \}^n$. (For completeness, we will show
in Appendix~\ref{app:lattice-proof} that $\realL$ is indeed a lattice;
we will also give an explicit generator matrix for $\realL$.)

Note that the lattice defined by $\realLc \triangleq \{ p^s \bfr : \bfr
\in \ZZ^n \}$ is a sublattice of $\realL$. Hence, we obtain a pair of
nested $\ZZ$-lattices $\realL \supseteq \realLc$ from the nested linear
codes $\{ \calC_i, 1 \le i \le s \}$.
	
Next, we lift this pair of nested $\ZZ$-lattices to a pair of nested
$\ZZ[i]$-lattices. That is, we set $\complexL = \realL + i \realL$ and
$\complexLc = \realLc+ i \realLc$. In this way, we obtain a pair of
nested $\ZZ[i]$-lattices $\complexL \supseteq \complexLc$. In
Appendix~\ref{app:relation}, we will show that there exist two generator
matrices $\bfG_{\complexL}$ and $\bfG_{\complexLc}$ satisfying
\begin{equation}\label{eq:lift_d}
	\bfG_{\complexLc} = {\rm diag}( \underbrace{p^s, \ldots, p^s}_{k_1},
	\underbrace{p^{s-1}, \ldots, p^{s-1}}_{k_2 - k_1}, \ldots,
	\underbrace{1, \ldots, 1}_{n - k_s}
	 )
	 \bfG_{\complexL}.
\end{equation}
It follows from Theorem~\ref{thm:construct} that
\[
	\complexL / \complexLc \cong
	(\ZZ[i]/ \langle p^s \rangle)^{k_1} \times \cdots \times
	 (\ZZ[i]/ \langle p \rangle )^{k_s - k_{s-1}}.
\]
In particular, the message rate $\Rmes = \frac{\sum_i k_i}{n} \log_2(
p^2 )$. When $s = 1$, this construction is reduced to the first
construction. Although this construction induces a more complicated
message space, it is able to produce pairs of nested lattices with
higher nominal coding gains, as shown in the following result.

\begin{prop}\label{prop:construction_d}
Let $\calC_1 \subseteq \cdots \subseteq \calC_s$ be nested linear codes
of length $n$ over $\ZZ/\langle p \rangle$ and let $\complexL \supseteq
\complexLc$ be a pair of nested lattices constructed from $\{\calC_i\}$.
Then $\gamma_c(\complexL / \complexLc)$ is lower bounded by
\[
\gamma_c(\complexL / \complexLc) \ge \frac{\min_{1\le i\le s} \{p^{2(i-1)}w_E^{\min}(\calC_i) \}}{p^{2(s - {\sum_{i = 1}^a k_i}/n)}},
\]
and $K(\complexL / \complexLc)$ is upper bounded by
\[
	K(\complexL / \complexLc) \le \begin{cases}
 2 \sum_{i = 1}^s 2^{A_i} A_i, & \text{when $p = 2$} \\
 2 \sum_{i = 1}^s A_i, & \text{when $p > 2$}
 \end{cases}
\]
where $A_i$ is the number of codewords in $\calC_i$ with minimum
Euclidean weight $w_E^{\min}(\calC_i)$.
\end{prop}
The proof is given in Appendix~\ref{app:construction-d}.

Now we will apply Propositions~\ref{prop:construction-a} and
\ref{prop:construction_d} to show the advantage of pairs of nested
lattices constructed via Construction~D. Let $\complexL_\text{A}
\supseteq \complexL'_\text{A}$ be a pair of nested lattices constructed
from a linear $[n, k]$ code $\calC$ (over $\ZZ/\langle p \rangle$) via
Construction~A. Then by Proposition~\ref{prop:construction-a},
$\gamma_c(\complexL_\text{A} / \complexL'_\text{A}) =
w_E^{\min}(\calC)/p^{2(1 - k/n)}$. Suppose that the linear code $\calC$
has an $[n, k']$ subcode $\calC'$ with $w_E^{\min}(\calC') \ge p^2
w_E^{\min}(\calC)$. Let $\complexL_\text{D} \supseteq
\complexL'_\text{D}$ be a pair of nested lattices constructed from
$\calC$ and $\calC'$ via Construction~D. Then by
Proposition~\ref{prop:construction_d},
\begin{align*}
\gamma_c(\complexL_\text{D} / \complexL'_\text{D}) &\ge \frac{p^2 w_E^{\min}(\calC)}{p^{2 (2 - (k + k')/n)}} \\
&= \frac{w_E^{\min}(\calC)}{p^{2 (1 - (k + k')/n)}} \\
&> \gamma_c(\complexL_\text{A} / \complexL'_\text{A}).
\end{align*}
In other words, given a pair of nested lattices via Construction~A,
there exists a pair of nested lattices via Construction~D with higher
nominal coding gain if the linear code $\calC$ has a subcode $\calC'$
with $w_E^{\min}(\calC') \ge p^2 w_E^{\min}(\calC)$.

\subsection{Design Examples}
We present three design examples to illustrate the design tools
developed in Sec.~\ref{sec:new_construct}. All of our design examples
feature short packet length and reasonable decoding complexity, since
the purpose of this paper is to demonstrate the potential of LNC schemes
in practical settings. (A more elaborate scheme, based on signal codes
\cite{SSF08}, is described in \cite{FSK11_isit}.)

\begin{exam}\label{exam:code-construction}
Consider a rate-$1/2$ terminated (feed-forward) convolutional code over
$\ZZ[i]/\langle 3 \rangle$ with $\nu$ memory elements. Suppose the input
sequence $u(D)$ is a polynomial of degree less than $\mu$. Then this
terminated convolutional code can be regarded as a $[2(\mu + \nu), \mu]$
linear block code $\calC$. Using the method based on complex
Construction~A, we obtain a pair of nested lattices $\complexL \supseteq
\complexLc$.

Note that the minimum Euclidean weight $w_E^{\min}(\calC)$ of $\calC$
can be bounded as
\[
	w_E^{\min}(\calC) \le 3(1 + \nu),
\]
for all rate-$1/2$ terminated (feed-forward) convolutional codes over
$\ZZ[i]/\langle 3 \rangle$. This upper bound can be verified by
considering the input sequence $u(D) = 1$. Hence, the nominal coding
gain $\gamma_c(\complexL / \complexLc)$ satisfies
\[
	\gamma_c(\complexL / \complexLc) \le 1 + \nu.
\]

When $\nu = 1, 2$ and $\mu \gg \nu$, this upper bound can be
asymptotically achieved by polynomial convolutional encoders shown in
Table~\ref{table:encoders}.

\begin{table}[t]
\caption{Polynomial convolutional encoders that
		asymptotically achieve the upper bound.}
		\centering
		\begin{tabular}{|c | c| c|}
			\hline
			$\nu$ & $\bfg(D)$ & $\gamma_c(\complexL / \complexLc)$ \\
			\hline
			$1$ & $[1 + (1+i)D, \ (1+i) + D]$ & 2 (3~dB) \\
			$2$ & $[1 + D + (1+i)D^2, \ (1+i) + (1-i)D +
D^2]$ & 3 (4.77~dB) \\
			\hline
		\end{tabular}
		\label{table:encoders}
\end{table}

Note that when $\nu = 1$ or $2$, the encoder state space size is $9$ or
$81$. Note also that the lattice decoder $\calD_{\complexL}$ can be
implemented through a modified Viterbi decoder as discussed in
Appendix~\ref{app:Viterbi}. Thus, this example demonstrates that a
nominal coding gain of $3$ to $5$~dB can be easily obtained with
reasonable decoding complexity.\hfill \IEEEQED
\end{exam}

Our next example illustrates how to use our design tools to improve an
existing construction presented in \cite{OE10}.
\begin{exam}\label{exam:lift_d}
Consider nested linear codes $\calC_1 \subseteq \calC_2$ of length $n$
over $\ZZ/\langle 2 \rangle$, where $\calC_1$ is an $[n, k_1, d_1]$ code
with $d_1 \ge 4$ and $\calC_2$ is the $[n, n]$ trivial code. Using the
method based on Construction~D, we obtain a pair of nested lattices
$\complexL \supseteq \complexLc$.

In this case, we will show that the nominal coding gain
$\gamma_c(\complexL / \complexLc) = 4/4^{(1-k_1/n)}$. On the one hand,
by Proposition~\ref{prop:construction_d},
\[
		\gamma_c(\complexL / \complexLc) \ge
		\frac{\min \{w_H^{\min}(\calC_1) , 4 w_H^{\min}(\calC_2) \}}{4^{(2 - {\sum_{i = 1}^2 k_i}/n)}}
		= 4/4^{(1-k_1/n)}.
\]
On the other hand, by definition,
\begin{align}
		\gamma_c(\complexL / \complexLc) &= d^2(\complexL / \complexLc)
		 / V(\complexLc)^{1/n} \nonumber \\
		&= d^2(\complexL / \complexLc) / 4^{(2 - {\sum_{i = 1}^2 k_i}/n)} \label{eq:VL} \\
		&\le 4 / 4^{(1-k_1/n)} \label{eq:dis}
\end{align}
where (\ref{eq:VL}) follows from the facts that $V(\complexLc) =
V(\complexL)4^{k_1+ k_2}$ and $V(\complexLc) = 4^{2n}$; (\ref{eq:dis})
follows from the fact that $(2, 0, \ldots, 0)$ is a lattice point in
$\complexL$ but not in $\complexLc$.
	
Finally, in Table~\ref{table:Hamming} we
list several candidates for $\calC_1$ as well as their
corresponding nominal coding gains. These candidates are all extended
Hamming codes with $d_1 = 4$. \hfill\IEEEQED
\end{exam}

We note that Ordentlich-Erez's construction in \cite{OE10} can be
regarded as a special case of Example~\ref{exam:lift_d}. In their
construction, $\calC_1$ is chosen as a rate $5/6$ cyclic LDPC code of
length $64800$. Example~\ref{exam:lift_d} suggests that their nominal
coding gain is $4/4^{1/6}$ ($5.02$ dB) with message rate $2(1 + 5/6)
\approx 3.67$. Example~\ref{exam:lift_d} also suggests that there are
many ways to improve the nominal coding gain. For example, when
$\calC_1$ is chosen as a $[256, 247]$ extended Hamming code, the nominal
coding gain is $5.81$ dB with message rate $2(1 + \frac{247}{256})
\approx 3.93$.

Our third example illustrates how to design high-coding-gain nested
lattices based on turbo lattices \cite{SSP10}.
\begin{exam}
Consider nested Turbo codes $\calC_1 \subseteq \calC_2$ over
$\ZZ/\langle 2 \rangle$. As shown in \cite{SSP10}, $\calC_1$ can be a
rate-$1/3$ Turbo code with $d_1 = 28$ and $\calC_2$ can be a rate-$1/2$
Turbo code with $d_2 = 13$. Using the method via Construction~D, we
obtain a pair of nested lattices $\complexL \supseteq \complexLc$. In
this case, by Proposition~\ref{prop:construction_d},
\[
		\gamma_c(\complexL / \complexLc) \ge
		\frac{\min \{d_1 , 4d_2 \}}{4^{(2 - {\sum_{i = 1}^2 k_i}/n)}} = 28/4^{(2 - 1/2 - 1/3)} = 7.45 \mbox{ dB}.
\]
The message rate is given by $\Rmes = 5/3 \approx 1.67$.
\end{exam}

Finally, some other design examples of high-performance nested lattice
codes, which are of a similar spirit, can be found, e.g., in
\cite{OZEGN11,FSK11_isit, TN11, TNBH12, QJ12}, Also, similar methods of
designing practical compute-and-forward have been recently proposed.
See, e.g., \cite{B11, HN11-submitted,HC12-submitted}.

\begin{table}[t]
\caption{Several extended Hamming codes and corresponding
nominal coding gains.}
\centering
\begin{tabular}{|c | c| c |}
\hline
$n$ & $k$ & $\gamma_c(\complexL / \complexLc)$ \\\hline
$32$ & $26$ & 3.08 (4.89~dB) \\
$64$ & $57$ & 3.44 (5.36~dB) \\
$128$ & $120$ & 3.67 (5.64~dB) \\
$256$ & $247$ & 3.81 (5.81~dB)\\\hline
\end{tabular}
\label{table:Hamming}
\end{table}

\section{Decoding Multiple Linear Combinations}\label{sec:choice}

In this section, we consider the problem when a receiver has the freedom
to choose coefficient vectors. For ease of presentation, we mainly focus
on the case of complex Construction~A in which the message space is a
vector space over $T / \langle \pi \rangle$. The main result of this
section is that, under separate decoding, the problem of decoding
multiple linear combinations is related to the \emph{shortest
independent vectors problem} \cite{Blomer00}, and can be solved through
some existing methods.

In general, upon deciding the coefficient vectors $\bfa_1, \ldots,
\bfa_m$, the receiver can perform joint decoding or separate decoding to
recover the linear combinations $\bfu_i = \bfa_i \bfW$. Here, we
confine our attention to separate decoding in which each linear
combination $\bfu_i = \bfa_i \bfW$ is decoded independently through the
use of $\calD(\bfy \mid \bfh, \bfa_i)$. In this case, the union bound
estimate on the decoding error for each $\bfa_i$ is
\[
 P_e(\bfh, \bfa_i) \lessapprox K(\Lambda
		/ \Lambda') \exp\left(- \frac{d^2(\Lambda / \Lambda')}
		{4N_0 \bfa_i \bfM \bfa_i\hh}\right).
\]

To optimize the above union bound estimates, the coefficient vectors
$\bfa_1, \ldots, \bfa_m$ should be chosen such that each $\bfa_i \bfM
\bfa_i\hh$ is made as small as possible under the constraint that
$\proj{\bfa}_1, \ldots, \proj{\bfa}_m$ are \emph{linearly independent}
over $T / \langle \pi \rangle$, where $\proj{\bfa}_i = \sigma(\bfa_i)$
is the natural projection of $\bfa_i$ (from $T$ to $T / \langle \pi
\rangle$). Clearly, this constraint ensures that every recovered linear
combination $\bfu_i$ is useful over $T / \langle \pi \rangle$.

 We say a solution $\{ \bfa_1, \ldots, \bfa_m \}$ is \emph{feasible} if
$\proj{\bfa}_1, \ldots, \proj{\bfa}_m$ are linearly independent over $T
/ \langle \pi \rangle$. Since each $\proj{\bfa}_i$ is of dimension $L$,
we assume that $m \le L$ because otherwise no feasible solution exists.

In the sequel, we will show that there exists a feasible solution that
\emph{simultaneously} optimizes each $\bfa_i \bfM \bfa_i\hh$. We call such
feasible solutions {\em dominant solutions}. Formally, let $\bfM = \bfL
\bfL^{\sf H}$ be the Cholesky decomposition of $\bfM$, where $\bfL$ is
some lower triangular matrix. (The existence of $\bfL$ comes from the
fact that $\bfM$ is Hermitian and positive-definite.) Clearly, $\bfa \bfM
\bfa\hh = \| \bfa \bfL \|^2$.
\begin{defi}[Dominant Solutions]
A feasible solution $\{ \bfa_1, \ldots, \bfa_m \}$ (with $\| \bfa_1 \bfL\| \le
\ldots \le \| \bfa_m \bfL\|$) is called a {\em dominant solution} if for
any feasible solution $\bfa_1', \ldots, \bfa_m'$ (with $\| \bfa_1' \bfL\|
\le \ldots \le \| \bfa_m' \bfL\|$), the following inequalities hold
\[
	\| \bfa_i \bfL \| \le \| \bfa_i' \bfL \|, \ i = 1, \ldots, m.
\]
\end{defi}

Although the dominant solutions seem to be a natural concept, the
existence of them is not immediate from the definition, and a separate
argument is needed.

\begin{thm}\label{thm:strongly}
A feasible solution $\{ \bfa_1, \ldots, \bfa_m\}$ defined by
	\begin{eqnarray*}
		\bfa_1 &=& \arg \min \left\{ \| \bfa \bfL \| \mid \proj{\bfa} \ \mbox{is nonzero} \right\} \\
		\bfa_2 &=& \arg \min \left\{ \| \bfa \bfL \| \mid \proj{\bfa}, \proj{\bfa}_1 \ \mbox{are linearly independent} \right\} \\
		&\vdots& \\
		\bfa_m &=& \arg \min \left \{ \| \bfa \bfL \| \mid \proj{\bfa}, \proj{\bfa}_1, \ldots, \proj{\bfa}_{m-1} \ \mbox{are linearly ind.}\right\}
	\end{eqnarray*}
	always exists, and is a dominant solution.
\end{thm}
The proof is given in Appendix~\ref{app:dominant}.

We now propose a three-step method of finding a dominant solution. In
the first step, we construct a ball $\calB(\rho) = \{ \bfx \in \CC^L \mid
\| \bfx \| \le \rho \}$ that contains $m$ lattice points $\bfv_1 \bfL,
\ldots, \bfv_m \bfL$ such that $\proj{\bfv}_1, \ldots, \proj{\bfv}_m$ are
linearly independent, where $\proj{\bfv}_i = \sigma(\bfv_i)$ is the
natural projection of $\bfv_i$. In the second step, we order all
lattice points within $\calB(\rho)$ based on their lengths, producing an
ordered set $\calS_\rho$ with $\|\bfv_1 \bfL\| \le \|\bfv_2 \bfL\| \le
\cdots \le \|\bfv_{|\calS_\rho|} \bfL \|$. Finally, we find a dominant
solution $\{ \bfa_1, \ldots, \bfa_m \}$ by using a greedy search algorithm
given as Algorithm~\ref{alg:reduction1}.

\begin{algorithm}[htp]
\caption{\textit{Greedy Search for Dominant Solution}}
\algsetup{linenosize=\footnotesize, linenodelimiter=.}
\emph{Input:} An ordered set $\calS_\rho = \{ \bfv_1 \bfL, \bfv_2 \bfL,
\ldots, \bfv_{|\calS_\rho|} \bfL \}$
with $\|\bfv_1 \bfL\| \le \|\bfv_2 \bfL\| \le \cdots
\le \|\bfv_{|\calS_\rho|} \bfL \|$.

\emph{Output:} An optimal solution $\{ \bfa_1, \ldots, \bfa_m \}$.

\begin{algorithmic}[1]
			\STATE{Set $\bfa_1 = \bfv_1$. Set $i = 1$ and $j = 1$.}
			\WHILE{$i < |\calS_b|$ and $j < m$}
			\STATE{Set $i = i+ 1$.}
			\IF{$\proj{\bfv}_i, \proj{\bfa}_1, \ldots, \proj{\bfa}_j$ are linearly independent}
			\STATE{Set $j = j+1$. Set $\bfa_j = \bfv_i$.}
			\ENDIF
			\ENDWHILE
		\end{algorithmic}
		\label{alg:reduction1}
\end{algorithm}

The correctness of our proposed method follows immediately from
Theorem~\ref{thm:strongly}. Our proposed method is in the spirit of
sphere-decoding algorithms, since sphere-decoding algorithms also
enumerate all lattice points within a ball centered at a given vector.
The selection of the radius $\rho$ plays an important role here, just as
it does for sphere-decoding algorithms. If $\rho$ is too large, then the
second step may incur excessive computations. If $\rho$ is too small,
then the first step may fail to construct a ball that contains $m$
linearly independent $\proj{\bfv}_1, \ldots, \proj{\bfv}_m$.

In practice, lattice-reduction algorithms \cite{Cassels71} may be used
to determine an appropriate radius $\rho$, as shown in the following
proposition.

\begin{prop}
Let $\{\bfb_1, \ldots, \bfb_L \}$ be a {\em reduced basis} \cite{Cassels71} for
$\bfL$. If $\rho$ is set to be $\|\bfb_m\|$, then the set $\calS_{\rho}$
contains at least $m$ lattice points $\bfv_1 \bfL, \ldots, \bfv_m \bfL$ such
that $\proj{\bfv}_1, \ldots, \proj{\bfv}_m$ are linearly independent.
\end{prop}
\begin{IEEEproof}
Let $\bfv_i = \bfb_i \bfL^{-1}$ for $i = 1, \ldots, L$. Let $\bfV$ be an $L
\times L$ matrix with $\bfv_i$ as its $i$th row. Since $\{\bfb_1, \ldots,
\bfb_L \}$ is a reduced basis, it follows that the matrix $\bfV$ is
invertible. In particular, $\proj{\bfv}_1, \ldots, \proj{\bfv}_{m}$ are
linearly independent for all integers $m \le L$.
\end{IEEEproof}

There are many existing lattice-reduction algorithms in the literature.
Among them, the Lenstra-Lenstra-Lov\'asz (LLL) algorithm \cite{LLL82} is
of particular importance. Moreover, the LLL algorithm has been extended
from real lattices to complex lattices over Euclidean domains
\cite{Napias96, GLM09}. Since $\ZZ[i]$ and $\ZZ[\omega]$ are special
cases of Euclidean domains, the extended LLL algorithm can be used to
handle the cases of $T = \ZZ[i]$ and $T = \ZZ[\omega]$.

Interestingly, when $L$ is small, some efficient lattice-reduction
algorithms can directly output dominant solutions. Such algorithms,
which are generalizations of Gauss' algorithm (see, e.g., \cite{Va91}),
are described in \cite{YW02, NS09}.

\section{Simulation Results}\label{sec:simulation}
As described in Section~\ref{sec:intro}, there are many potential
application scenarios for LNC, the most promising of which may involve
multicasting from one (or more) sources to multiple destinations via a
wireless relay network. Since we wish to avoid introducing higher-layer
issues (e.g., scheduling), in this paper, we focus here on a
two-transmitter, single receiver multiple-access configuration, which
may be regarded as a building block component of a more complicated and
realistic network application. In particular, we focus on the following
three scenarios:
\begin{enumerate}
\item The channel gains are fixed; the receiver chooses a single linear
function.
\item The channel gains are Rayleigh faded; the receiver chooses a
single linear function.
\item The channel gains are Rayleigh faded; the receiver chooses two
linear functions.
\end{enumerate}
In each scenario, we evaluate the performance of four LNC schemes:
the Nazer-Gastpar scheme, two LNC schemes
proposed in Example~\ref{exam:code-construction}, and the baseline LNC
scheme over $\ZZ[i]/\langle 3 \rangle$ as defined in
Sec.~\ref{sec:design}. Since we are interested in LNC schemes with
short packet lengths, each transmitted signal consists of $200$ complex
symbols in our simulations.

\begin{figure}[t]
\ifCLASSOPTIONonecolumn
\centering\includegraphics[width=0.5\columnwidth]{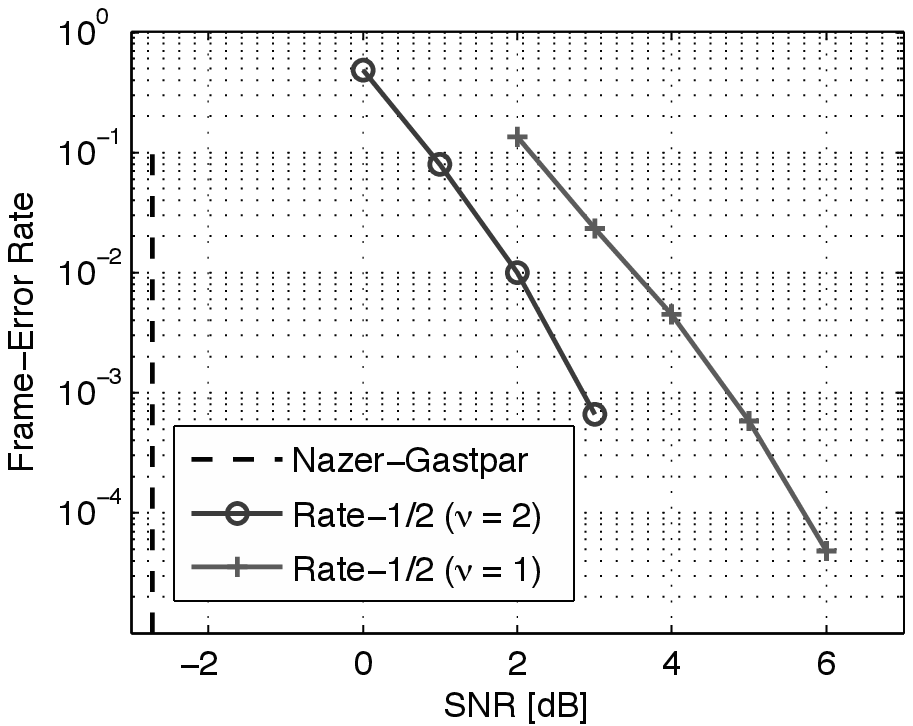}
\else
\centering\includegraphics[width=0.9\columnwidth]{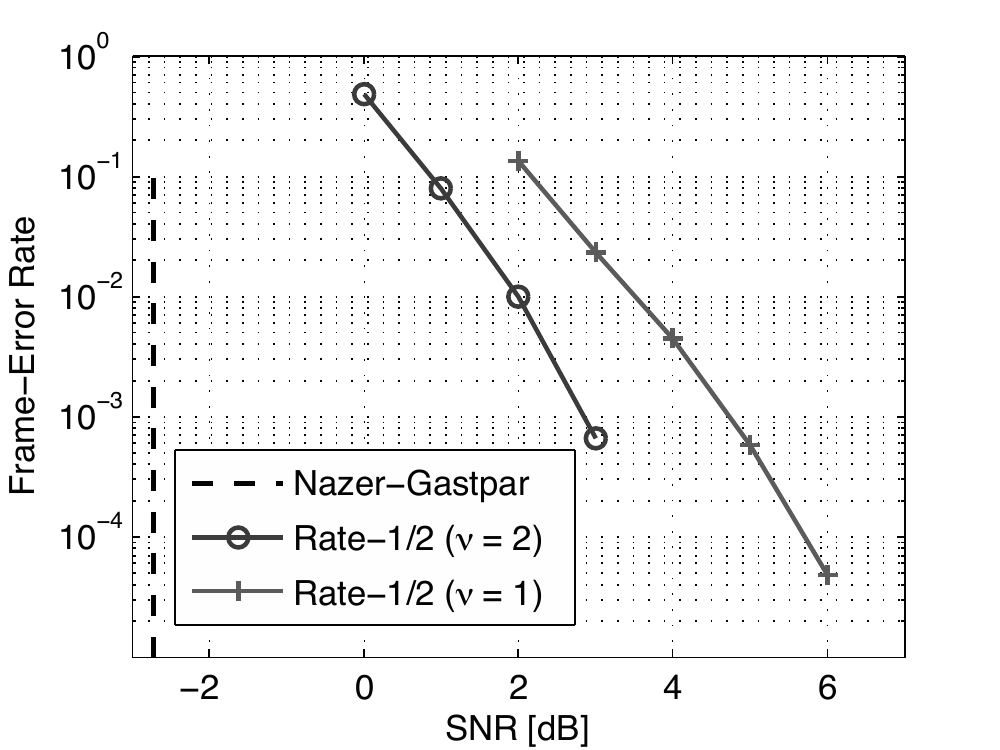}
\fi
\caption{Error performance of three LNC schemes in Scenario 1.}
\label{fig:nSENR}
\end{figure}

\subsection{Scenario~1 (Fixed Channel Gains; Single Coefficient Vector)}
Fig.~\ref{fig:nSENR} depicts the frame-error rates of three LNC schemes
as a function of $\SNR$. Here, the channel-gain vector $\bfh$ is set to
$\bfh = [-1.17 + 2.15i \ 1.25 - 1.63i]$. Nevertheless, as we have shown
in Sec.~\ref{sec:design}, the results are not particularly sensitive to
the choice for $\bfh$; similar results are achieved for other fixed
choices for $\bfh$. For the two LNC schemes proposed in
Example~\ref{exam:code-construction}, the parameter $\mu + \nu$ is set
to $100$ and the corresponding message rates are
$\frac{99}{100}\log_2(3)$ ($\nu = 1$) and $\frac{98}{100}\log_2(3)$
($\nu = 2$), respectively. For the Nazer-Gastpar scheme, the message
rate is set to $\log_2(3)$, which is quite close to the previous two
message rates. The decoding rule for the Nazer-Gastpar scheme is as
follows: a frame error occurs if and only if $\log_2(3) \ge
\log_2(\SNR/\bfa \bfM \bfa\hh)$, where $\bfa$ is the single coefficient
vector. From Fig.~\ref{fig:nSENR}, we observe that the gap to the
Nazer-Gastpar scheme is around $5$~dB at an error-rate of $1\%$. We
also observe that the second LNC scheme (with state space of size $81$)
outperforms the first LNC scheme (with state space of size $9$) by about
$2$~dB.

\begin{figure}[t]
\centering
\ifCLASSOPTIONonecolumn
\begin{tabular}{c}
\includegraphics[width=0.4\columnwidth]{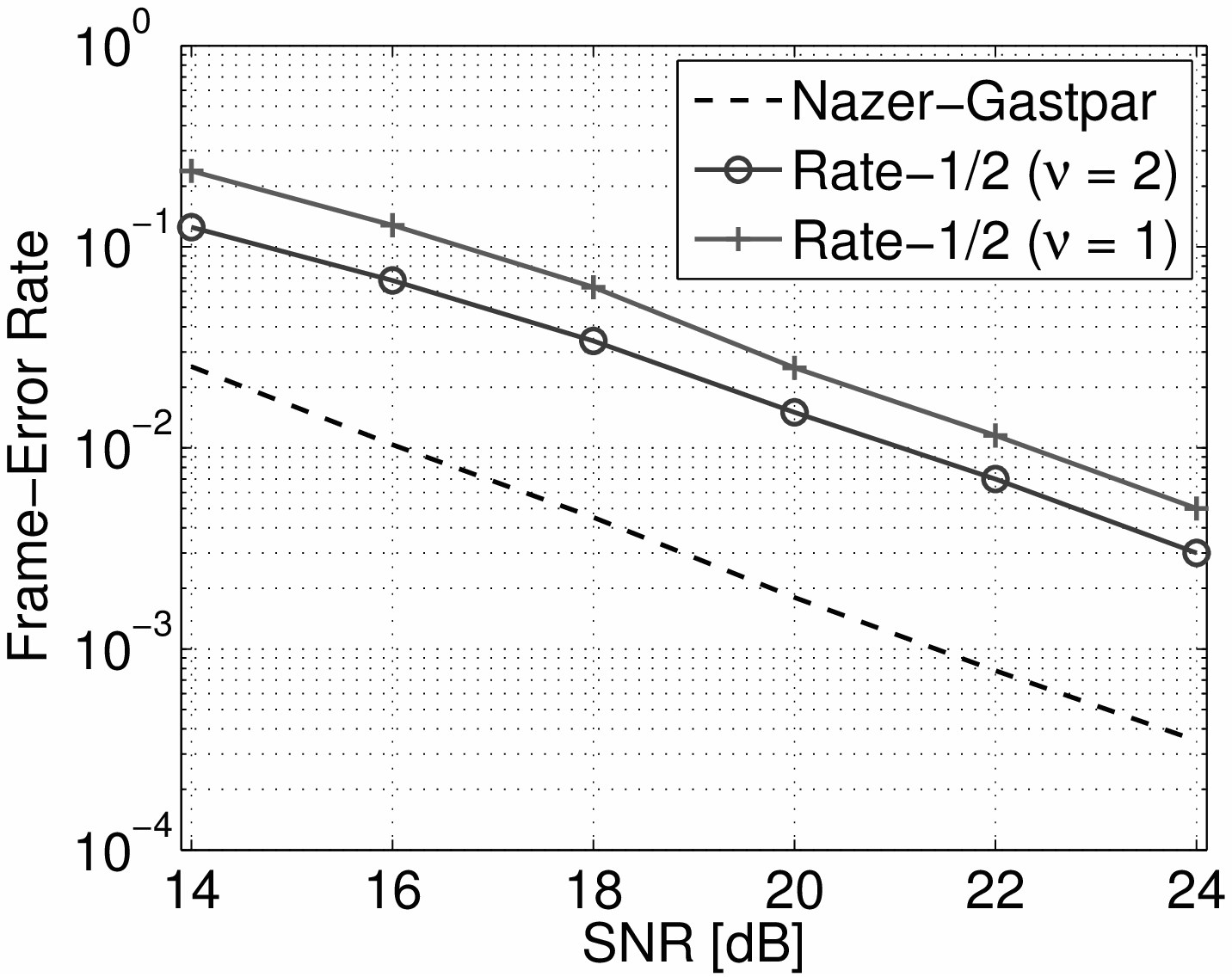}\\{\small (a)}\\
\includegraphics[width=0.4\columnwidth]{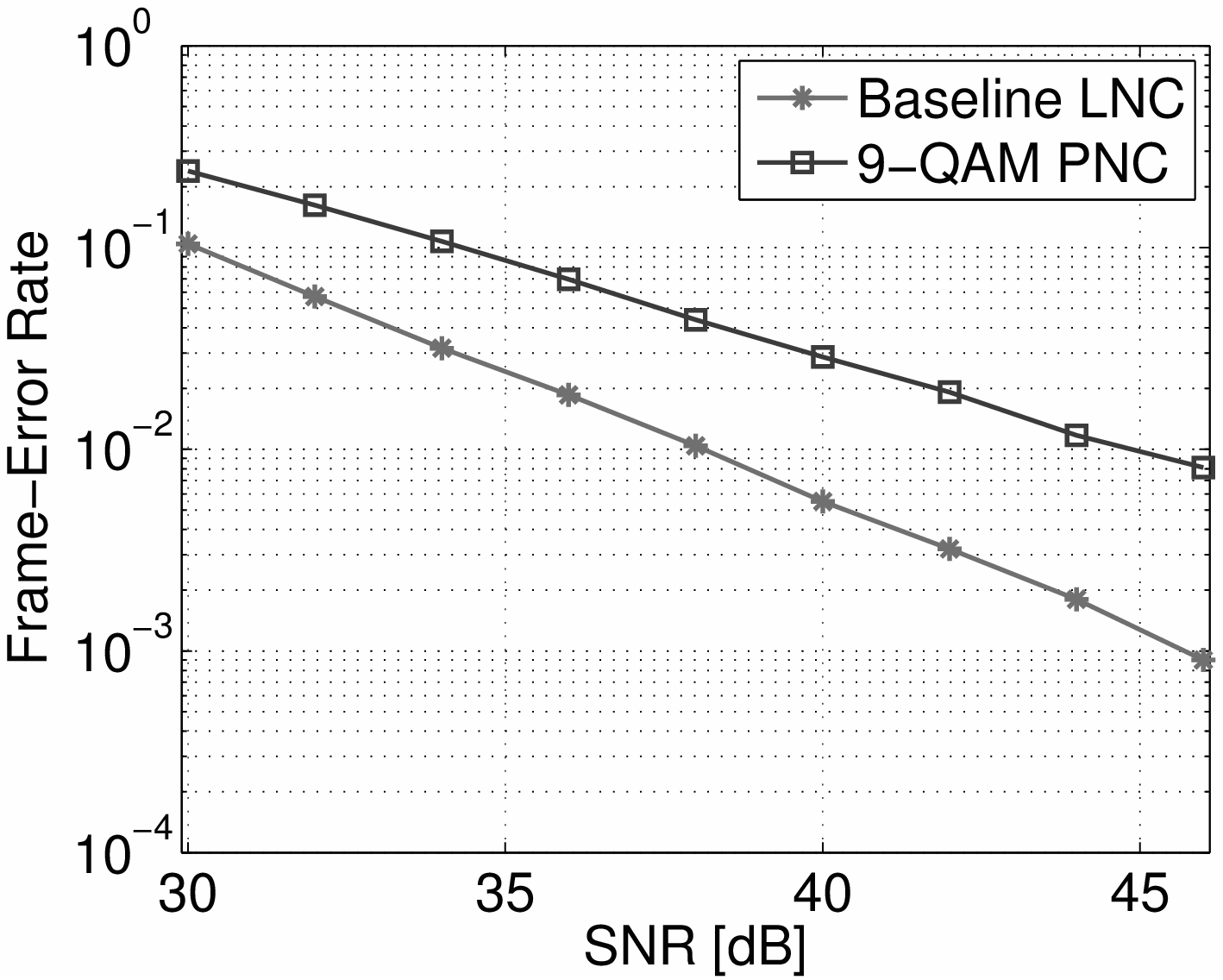}\\{\small (b)}
\end{tabular}
\else
\begin{tabular}{c}
\includegraphics[width=0.83\columnwidth]{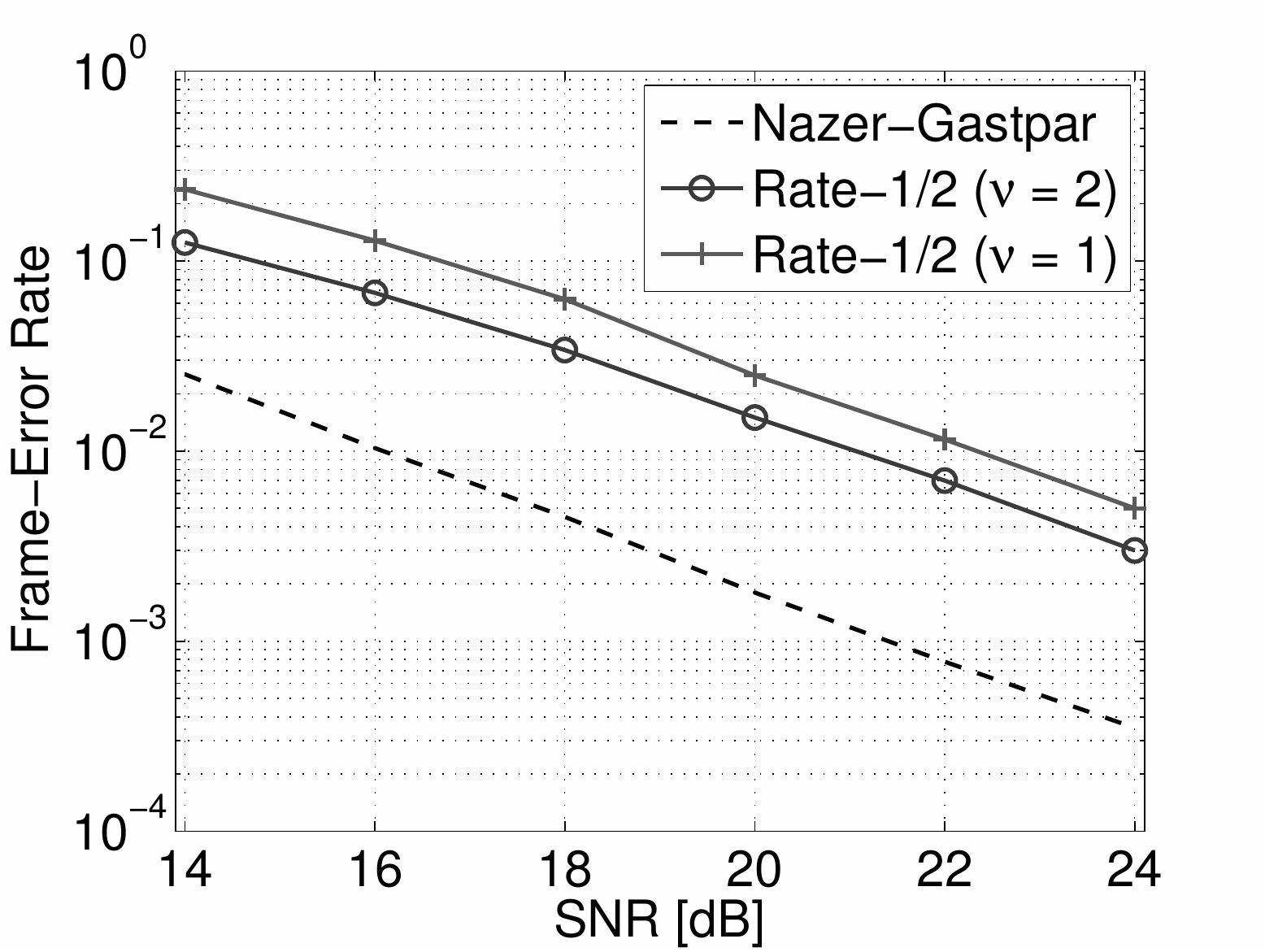}\\{\small (a)}\\
\includegraphics[width=0.83\columnwidth]{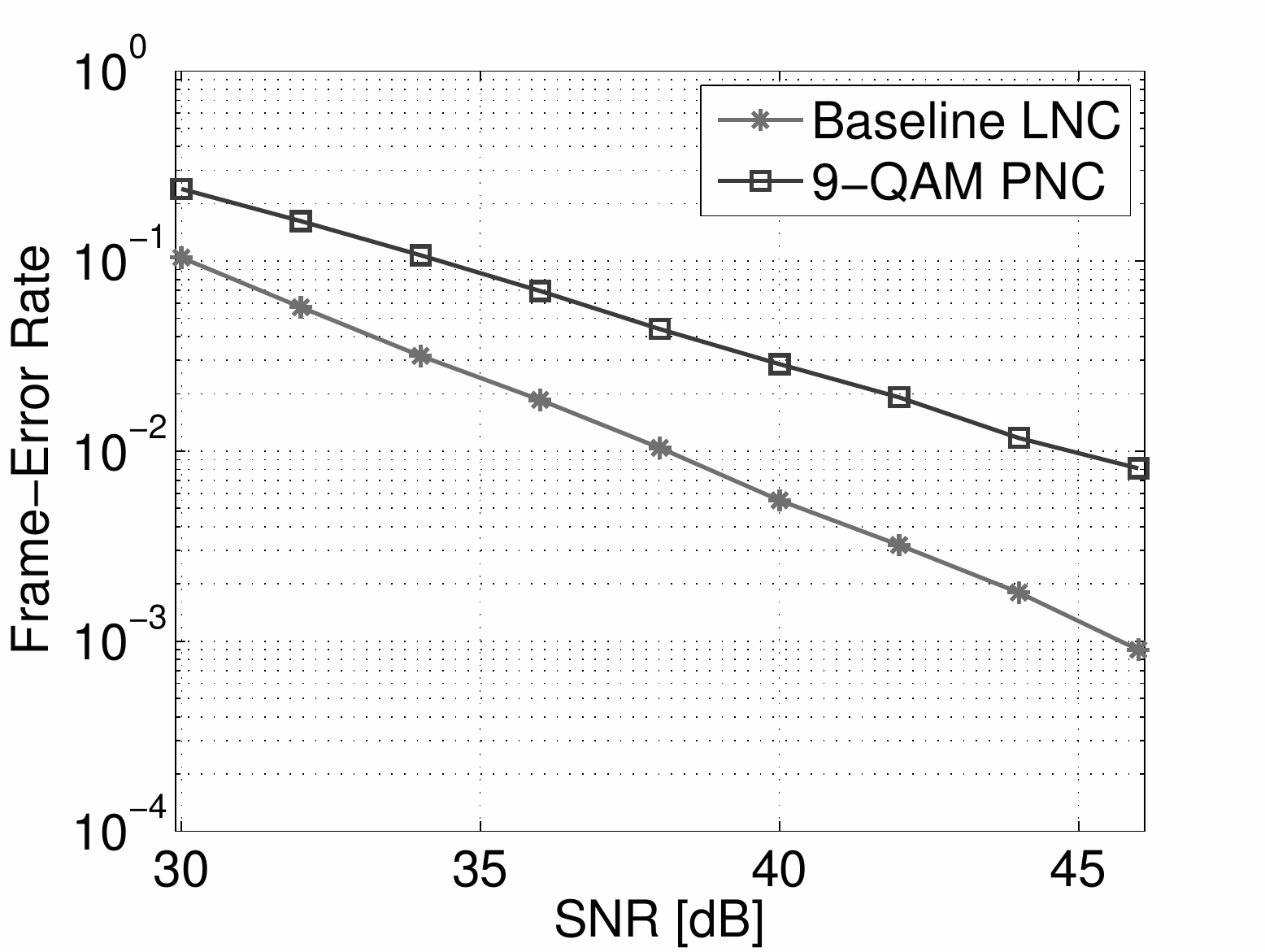}\\{\small (b)}
\end{tabular}
\fi
\caption{Error performance of various LNC schemes in Scenario 2.}
\label{fig:SNR}
\end{figure}

\subsection{Scenario~2 (Rayleigh-faded Channel Gains; Single Coefficient Vector)}
Fig.~\ref{fig:SNR}(a) shows the frame-error rates of three LNC schemes as a
function of $\SNR$. The setup is the same as in Scenario~1, except that
the coefficient vector $\bfa$ changes with $\bfh$. As seen in
Fig.~\ref{fig:SNR}(a), the gap to the Nazer-Gastpar scheme is around $5$~dB
at an error-rate of $1\%$.

Fig.~\ref{fig:SNR}(b) shows the frame-error rates of the baseline LNC
scheme (over $\ZZ[i]^{200}/3\ZZ[i]^{200}$) and the $9$-QAM PNC scheme
described in Example~\ref{exam:QAM}. For the $9$-QAM scheme, the
coefficient vector $\bfa$ is set to $[1 \ 1]$ as explained in
Example~\ref{exam:QAM}. To make a fair comparison, the coefficient
vector $\bfa$ in the baseline LNC scheme satisfies $a_1 \ne 0, a_2 \ne
0$, which comes from the ``exclusive law of network coding'' as
discussed in \cite{PY07, APT08}. As seen in Fig.~\ref{fig:SNR}(b), the
baseline LNC scheme outperforms the $9$-QAM scheme by more than $6$~dB
at an error-rate of $1\%$. In other words, even the baseline LNC scheme
is able to effectively mitigate phase misalignment due to Rayleigh
fading. Finally, note that Fig.~\ref{fig:SNR}(a) and Fig.~\ref{fig:SNR}(b)
are separated because they have different message rates ($\log_2(3)$ in
Fig.~\ref{fig:SNR}(a) and $2\log_2(3)$ in Fig.~\ref{fig:SNR}(b)).

\begin{figure}[t]
\ifCLASSOPTIONonecolumn
\centering\includegraphics[width=0.45\columnwidth]{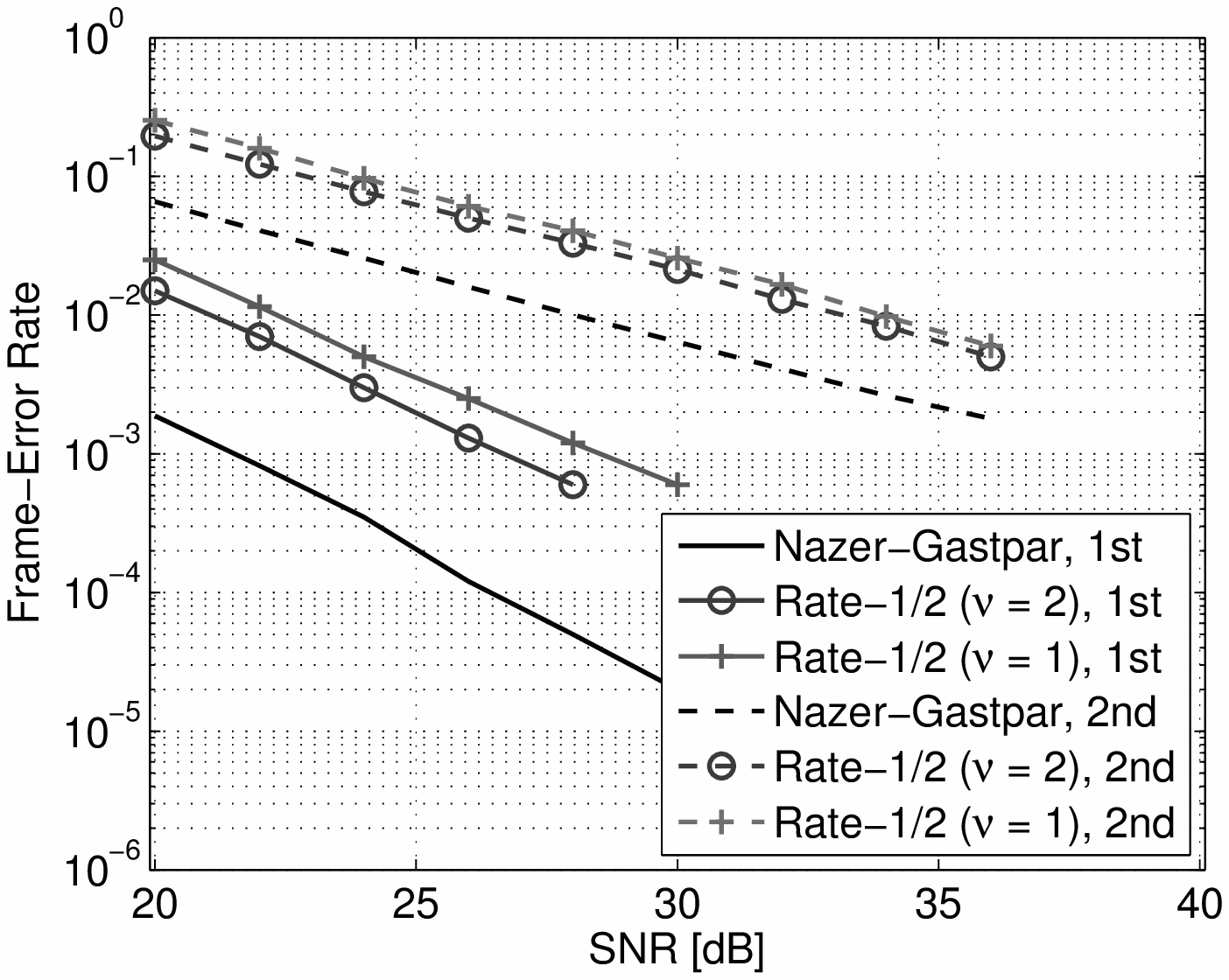}
\else
\centering\includegraphics[width=0.9\columnwidth]{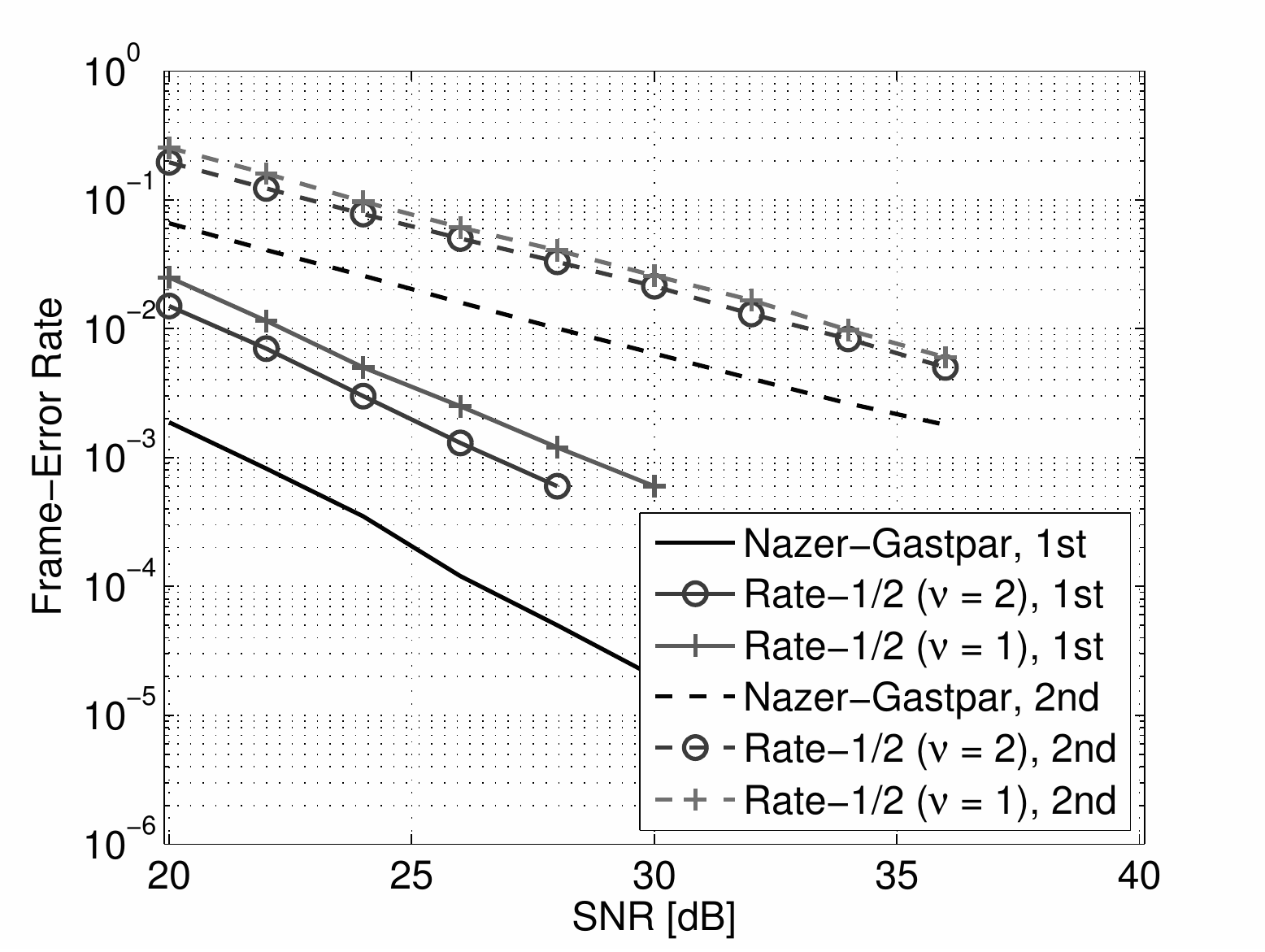}
\fi
\caption{Error performance of three LNC schemes in Scenario 3.}
\label{fig:SNR2}
\end{figure}

\subsection{Scenario~3 (Rayleigh-faded Channel Gains; Two Coefficient Vectors)}
Fig.~\ref{fig:SNR2} depicts the frame-error rates of three LNC schemes
as a function of $\SNR$. Here the two coefficient vectors are chosen by
using the lattice-reduction algorithm proposed in \cite{YW02}. The
configurations for the three LNC schemes are precisely the same as those
in Fig.~\ref{fig:SNR}. The frame-error rates for the first linear
combination are depicted in solid lines, while the error rates for the
second linear combination are depicted in dashed lines. From
Fig.~\ref{fig:SNR2}, we observe similar trends of error rates as in
Fig.~\ref{fig:SNR}. We also observe that the first linear combination is
much more reliable than the second one.

\section{Conclusion}\label{sec:conclusion}

In this paper, the problem of constructing LNC schemes via
finite-dimensional nested lattices has been studied. A generic LNC
scheme has been defined based on an arbitrary pair of nested lattices.
The message space of the generic scheme is a finite module in general,
whose structure may be analyzed using the Smith normal form theorem.
These results not only give rise to a convenient characterization of the
message space of the Nazer-Gastpar scheme, but also lead to several
generalized constructions of LNC schemes. All of these constructions are
compatible with header-based random linear network coding.

An estimate of the error probability for hypercube-shaped LNC
schemes has been derived, showing that the pair of nested lattices
$\Lambda \supseteq \Lambda'$ should be designed such that $d(\Lambda /
\Lambda')$ is maximized and $K(\Lambda / \Lambda')$ is minimized. These
criteria lead to several specific methods for optimizing nested
lattices. In particular, the nominal coding gain for pairs of nested
lattices has been introduced, which serves as an important figure of
merit for comparing various LNC schemes. In addition, several concrete
examples of practical LNC schemes have been provided, showing that a
nominal coding gain of $3$ to $7.5$~dB is easily obtained under
reasonable decoding complexity and short packet length. Finally, the
problem of choosing multiple coefficient vectors is discussed, which is
connected to some well-studied lattice problems, such as the shortest
independent vectors problem and the lattice reduction problem.

We believe that there is still much work to be done in this area. One
direction for follow-up work would be the design and analysis of
higher-layer scheduling algorithms for LNC schemes. Another direction
would be the study of more general shaping methods beyond hypercube
shaping. A particular example along this direction is given in
\cite{QJ12}. A third direction would be the construction of more
powerful LNC schemes, which has been partially explored in several
recent papers, e.g., \cite{OZEGN11,FSK11_isit, TN11, TNBH12}. We
believe that the algebraic framework given in this paper can serve as a
good basis for these developments.

\appendix

\subsection{Proof of Theorem~\ref{thm:criterion}}\label{app:error}

We upper bound the error probability
\mbox{$\Pr[\mathcal{Q}_{\Lambda}^{\nn}(\bfn) \notin \Lambda']$}.
{Consider the (non-lattice) set $\{ \Lambda \setminus \Lambda'\} \cup \{
\mathbf{0} \}$, i.e., the set difference $\Lambda \setminus \Lambda'$
adjoined with the zero vector. Let $\calR_V(\mathbf{0})$ be the Voronoi
region of $\mathbf{0}$ in the set $\{ \Lambda \setminus \Lambda'\} \cup
\{ \mathbf{0} \}$, i.e.,}
\[
\calR_V(\mathbf{0}) = \left\{\bfx \in \CC^n:
 \forall \bflambda \in \Lambda \setminus \Lambda'
 \left(\|\bfx - \mathbf{0}\| \le \|\bfx - \bflambda\|\right)
 \right\}.
\]
We have the following upper bound for
$\Pr[\mathcal{Q}_{\Lambda}^{\nn}(\bfn) \notin \Lambda']$.

\begin{lem}
	$\Pr[\mathcal{Q}_{\Lambda}^{\nn}(\bfn) \notin \Lambda'] \le \Pr[\bfn \notin \calR_V(\mathbf{0})]$.
\end{lem}
\begin{IEEEproof}
\begin{align*}
\Pr[\bfn \in \calR_V(\mathbf{0})]
&= \Pr[ \forall \bflambda \in \Lambda \setminus \Lambda'
\left(\|\bfn - \mathbf{0}\| \le \|\bfn - \bflambda\|\right)
] \\
&= \Pr[ \forall \bflambda \in \Lambda \setminus \Lambda'
\left(
\|\bfn - \mathbf{0}\| < \|\bfn - \bflambda\| \right) ].
\end{align*}
	Note that if $\|\bfn - \mathbf{0}\| <
\|\bfn - \bflambda\|$ for all $\bflambda \in \Lambda \setminus
\Lambda'$, then $\mathcal{Q}_{\Lambda}^{\nn}(\bfn) \notin \Lambda
\setminus \Lambda'$, as $\bfzero$ is closer to $\bfn$
than any element in $\Lambda \setminus \Lambda'$. Thus,
	\[
	\Pr[\bfn \in \calR_V(\mathbf{0})] \le \Pr[\mathcal{Q}_{\Lambda}^{\nn}(\bfn) \notin \Lambda \setminus \Lambda'] = \Pr[\mathcal{Q}_{\Lambda}^{\nn}(\bfn) \in \Lambda'].
	\]
\end{IEEEproof}

We further upper bound the probability $\Pr[\bfn \notin
\calR_V(\mathbf{0})]$. Let $\textrm{Nbr}(\Lambda \setminus \Lambda')
\subseteq \Lambda \setminus \Lambda'$ denote the set of neighbors of
$\mathbf{0}$ in $\Lambda \setminus \Lambda'$, i.e.,
$\textrm{Nbr}(\Lambda \setminus \Lambda')$ is the smallest subset of
$\Lambda \setminus \Lambda'$ such that $\calR_V(\mathbf{0})$ is
precisely the set
\[
\left\{\bfx \in \CC^n:
\forall \bflambda \in
\textrm{Nbr}(\Lambda \setminus \Lambda')
\left(
\|\bfx - \mathbf{0}\| \le \|\bfx - \bflambda\|\right)
\right\}.
\]

Then, for any $\nu>0$, we have
\begin{align}
&\ P[\bfn \not\in \calR_V(\mathbf{0})] \nonumber \\
&= P\left[\|\bfn\|^2 \geq \|\bfn - \bflambda\|^2,\; \text{some $\bflambda \in \textrm{Nbr}(\Lambda \setminus \Lambda')$}\right] \nonumber \\
&= P\left[\Re\{\bflambda^{\sf H} \bfn\} \geq \|\bflambda\|^2/2,\; \text{some $\bflambda \in \textrm{Nbr}(\Lambda \setminus \Lambda')$}\right] \nonumber \\
&\leq \sum_{\bflambda \in \textrm{Nbr}(\Lambda \setminus \Lambda')} P\left[\Re\{\bflambda^{\sf H} \bfn\} \geq \|\bflambda\|^2/2\right] \label{eq:perr1} \\
&\leq \sum_{\bflambda \in \textrm{Nbr}(\Lambda \setminus \Lambda')} \exp({-\nu \|\bflambda\|^2/2}) E\left[\exp({\nu \Re\{\bflambda^{\sf H} \bfn\}})\right], \label{eq:perr2}
\end{align}
where (\ref{eq:perr1}) follows from the union bound and (\ref{eq:perr2})
follows from the Chernoff bound. Since $\bfn = \sum_\ell (\alpha h_\ell
- a_\ell) \bfx_\ell + \alpha \bfz$, we have
\begin{align}
&\ E\left[\exp\big({\nu \Re\{\bflambda^{\sf H} \bfn\}}\big)\right] \nonumber \\
&= E\left[\exp\left({\nu \Re\left\{\bflambda^{\sf H} \left(\sum_\ell (\alpha h_\ell - a_\ell)\bfx_\ell + \alpha \bfz\right) \right\}}\right)\right] \nonumber \\
&= E\left[\exp({\nu \Re\{\bflambda^{\sf H} \alpha \bfz\}})\right] \nonumber \\
&\quad \ \cdot \prod_\ell E\left[\exp({\nu \Re\{\bflambda^{\sf H} (\alpha h_\ell - a_\ell)\bfx_\ell \}})\right] \label{eq:perr3} \\
&= \exp\left({\frac{1}{4}\nu^2 \|\bflambda\|^2 |\alpha|^2 N_0} \right) \nonumber \\
&\quad \ \cdot \prod_\ell E\left[\exp({\nu \Re\{\bflambda^{\sf H} (\alpha h_\ell - a_\ell)\bfx_\ell \}})\right] \label{eq:perr4}
\end{align}
where (\ref{eq:perr3}) follows from the independence of
$\bfx_1,\ldots,\bfx_L,\bfz$ and (\ref{eq:perr4}) follows from the
moment-generating function of a circularly symmetric complex Gaussian
random vector.

\begin{lem}
Let $\bfx \in \CC^n$ be a complex random vector uniformly distributed
over a hypercube $\gamma \bfU \calH_n$ for some $\gamma >0$ and some $n
\times n$ unitary matrix. Then
\[
E\left[\exp({\Re\{\bfv^{\sf H} \bfx\}})\right] \leq \exp({\|\bfv\|^2
\gamma^2/24}).
\]
\end{lem}
\begin{IEEEproof}
 First, we consider a special case where the unitary matrix $\bfU =
\bfI_n$. In this case, we have
 \begin{align}
 &\ E\left[\exp({\Re\{\bfv^{\sf H} \bfx\}})\right] \nonumber \\
 &= E\left[\exp({\Re\{\bfv\}^T\Re\{\bfx\} + \Im\{\bfv\}^T\Im\{\bfx\}})\right] \nonumber \\
 &= E\left[\exp\left({\sum_{i=1}^n \left(\Re\{\bfv_i\}\Re\{\bfx_i\} + \Im\{\bfv_i\}\Im\{\bfx_i\}\right)}\right)\right] \nonumber \\
 &= \prod_{i=1}^n E\left[\exp({\Re\{\bfv_i\}\Re\{\bfx_i\}}\right] E\left[\exp{\Im\{\bfv_i\}\Im\{\bfx_i\}})\right] \label{eq:plem-1} \\
 &= \prod_{i=1}^n \frac{\sinh(\Re\{\bfv_i\}\gamma/2)}{\Re\{\bfv_i\}\gamma/2} \frac{\sinh(\Im\{\bfv_i\}\gamma/2)}{\Im\{\bfv_i\}\gamma/2} \label{eq:plem-2} \\
 &\leq \prod_{i=1}^n \exp\left(\frac{(\Re\{\bfv_i\}\gamma)^2}{24}\right) \exp\left(\frac{(\Im\{\bfv_i\}\gamma)^2}{24}\right) \label{eq:plem-3} \\
 &= \exp\left( \frac{\gamma^2}{24}\|\bfv\|^2 \right) \nonumber
 \end{align}
where (\ref{eq:plem-1}) follows from the independence among each
real/imaginary component, (\ref{eq:plem-2}) follows from the
moment-generating function of a uniform random variable (note that both
$\Re\{\bfx_i\}$ and $\Im\{\bfx_i\}$ are uniformly distributed over
$[-\gamma/2, \gamma/2]$), and (\ref{eq:plem-3}) follows from $\sinh(x)/x
\leq \exp({x^2/6})$ (which can be obtained by simple Taylor expansion).

Then we consider a general unitary matrix $\bfU$. In this case, we have
$\bfx = \bfU \bfx'$, where $\bfx' \in \gamma [-1/2, 1/2]^{2n}$, i.e., both
$\Re\{\bfx_i'\}$ and $\Im\{\bfx_i'\}$ are uniformly distributed over
$[-\gamma/2, \gamma/2]$. Hence,
\begin{align*}
 E\left[\exp({\Re\{\bfv^{\sf H} \bfx\}})\right]
 &= E\left[\exp({\Re\{\bfv^{\sf H} \bfU \bfx'\}})\right]  \\
 &= E\left[\exp({\Re\{(\bfU^{\sf H}\bfv)^{\sf H} \bfx'\}})\right]  \\
 &\leq \exp\left( \frac{\gamma^2}{24}\|\bfU^{\sf H} \bfv\|^2 \right)  \\
 &= \exp\left( \frac{\gamma^2}{24}\|\bfv\|^2 \right).
\end{align*}
\end{IEEEproof}

Note that $P = \frac{1}{n} E[\| \bfx_\ell \|^2] = \gamma^2/6$. Thus, we
have
\begin{align*}
&\ E\left[\exp({\nu \Re\{\bflambda^{\sf H} \bfn\}})\right]  \\
&\leq \exp\left({\frac{1}{4}\nu^2 \|\bflambda\|^2 |\alpha|^2 N_0} \right) \prod_\ell \exp({\|\nu \bflambda (\alpha h_\ell - a_\ell)\|^2 P/4})  \\
&= \exp\left({\frac{1}{4}\nu^2 \|\bflambda\|^2 |\alpha|^2 N_0 + \|\nu \bflambda \|^2 \|\alpha \bfh - \bfa\|^2 P/4} \right)  \\
&= \exp\left({\frac{1}{4} \|\bflambda\|^2 \nu^2 N_0 Q(\bfa,\alpha)}\right),
\end{align*}
where the quantity $Q(\bfa,\alpha)$ is given by
\[
Q(\bfa,\alpha) = |\alpha|^2 + \SNR \|\alpha \bfh - \bfa\|^2
\]
and $\SNR = P/N_0$.

It follows that, for all $\nu > 0$,
\begin{align*}
&\ \Pr[\bfn \not\in \calR_V(\mathbf{0})] \\
&\leq \sum_{\bflambda \in \textrm{Nbr}(\Lambda \setminus \Lambda')} \exp\left({-\nu \|\bflambda\|^2/2 + \frac{1}{4} \|\bflambda\|^2 \nu^2 N_0 Q(\bfa,\alpha)}\right).
\end{align*}
Choosing $\nu = 1/(N_0 Q(\bfa,\alpha))$, we have
\begin{align*}
\Pr[\bfn \not\in \calR_V(\mathbf{0})] &\le \sum_{\bflambda \in \textrm{Nbr}(\Lambda \setminus \Lambda')} \exp\left({- \frac{\|\bflambda\|^2}{4N_0 Q(\bfa,\alpha)}} \right) \\
& \approx K(\Lambda / \Lambda') \exp\left(- \frac{d^2(\Lambda / \Lambda')}{4N_0Q(\bfa,\alpha)}\right)
\end{align*}
for high signal-to-noise ratios. Therefore, we have
\begin{align*}
	\Pr[\mathcal{Q}_{\Lambda}^{\nn}(\bfn) \notin \Lambda']
	&\le \Pr[\bfn \notin \calR_V(\mathbf{0})]  \\
	&\lessapprox K(\Lambda / \Lambda') \exp\left(- \frac{d^2(\Lambda / \Lambda')}{4N_0Q(\bfa,\alpha)}\right).
\end{align*}
Since $\alpha$ can be carefully chosen, we have
\[
	\Pr[\mathcal{Q}_{\Lambda}^{\nn}(\bfn) \notin \Lambda'] \lessapprox \min_{\alpha \in \CC}
	K(\Lambda / \Lambda') \exp\left(- \frac{d^2(\Lambda / \Lambda')}{4N_0Q(\bfa,\alpha)}\right),
\]
completing the proof for the first part of Theorem~\ref{thm:criterion}.
The second part of Theorem~\ref{thm:criterion} follows immediately when
the optimal value of $\alpha$ is substituted.

\subsection{Proof of Proposition~\ref{prop:construction-a}}\label{app:construction-a}

Recall that $d(\realL / \realLc)$ is the length of the shortest vectors
in the set difference $\realL \setminus \realLc$. Hence, we have
\[
d(\realL / \realLc) = \min_{\bfc \ne \bfzero} \| \best{\bfc} \|;
\]
equivalently, $d^2(\realL / \realLc) = \min_{\bfc \ne \bfzero} \|
\best{\bfc} \|^2 = w_E^{\min}(\calC)$. Recall that $\complexL = \realL +
i \realL$. That is, $\complexL = \realL \times \realL$. Hence, we have
\[
	d^2(\complexL / \complexLc) =
	d^2(\realL / \realLc) = w_E^{\min}(\calC).
\]
 Note that $V(\complexLc) = p^{2n}$ and $V(\complexLc) / V(\complexL) =
p^{2k}$. Hence, we have $V(\complexL) = p^{2(n-k)}$. Combining the
above two results, we have
\[
	\gamma_c(\complexL / \complexLc) = w_E^{\min}(\calC) / p^{2(1 - k/n)}.
\]

We then turn to $K(\realL / \realLc)$ and $K(\complexL / \complexLc)$.
When $p = 2$, the minimum Euclidean weight $w_E^{\min}(\calC)$ of
$\calC$ is precisely the minimum Hamming weight of $\calC$. In this
case, $K(\realL / \realLc) =
\left(w_E^{\min}(\calC)\right)2^{w_E^{\min}(\calC)}$, as shown in
\cite{CS99}. When $p > 2$, the set different $\realL \setminus \realLc$
can be expressed as
\[
	\realL \setminus \realLc = \bigcup_{\bfc \ne \bfzero} \left\{ \best{\bfc} + \realLc \right\}.
\]
In this case, $\best{\bfc}$ is the unique coset leader for the coset
$\best{\bfc} + \realLc$. Thus, the number $K(\realL / \realLc)$ of the
shortest vectors in $\realL \setminus \realLc$ is precisely the number
$A\left(w_E^{\min}(\calC)\right)$ of coset leaders with $\| \best{\bfc}
\|^2 = w_E^{\min}(\calC)$. Hence, we have
\[
	K(\realL / \realLc) = \begin{cases}
 	A\left(w_E^{\min}(\calC)\right)2^{w_E^{\min}(\calC)}, & \text{when $p = 2$}, \\
 	A\left(w_E^{\min}(\calC)\right), & \text{when $p > 2$}.
	\end{cases}
\]
Recall that $\complexLc = \realLc + i \realLc$. That is, $\complexLc =
\realLc \times \realLc$. It follows that $K(\complexL / \complexLc) =
2K(\realL / \realLc)$, completing the proof.

\subsection{Proof of Proposition~\ref{prop:construction-aa}}\label{app:construction-aa}

The proof is analogous to that of Proposition~\ref{prop:construction-a}
with two differences. First, $p$ is replaced by $| \pi |$ in the
expression of $\gamma_c(\complexL / \complexLc)$. This difference comes
from the fact that $V(\complexLc) = | \pi |^{2n}$ and $V(\complexLc) /
V(\complexL) = |\pi|^{2k}$. Second, the case of $| \pi | = 2$ gives an
expression of $A\left(w_E^{\min}(\calC)\right)4^{w_E^{\min}(\calC)}$ for
$K(\complexL /\complexLc)$. This is because if the coset $\bfc +
\complexLc$ contains one shortest vector in $\complexL \setminus
\complexLc$, then a total of $4^{w_E^{\min}(\calC)}$ shortest vectors
can be found in the coset $\bfc + \complexLc$. Suppose that $(c_1,
\ldots, c_n)$ is one such shortest vector in $\bfc + \complexLc$. Then,
$(c_1, \ldots, c_n)$ has precisely $w_E^{\min}(\calC)$ nonzero elements.
Moreover, for each nonzero element, say $c_j$, if we change it to one of
$\{ - c_j, i \times c_j, (-i) \times c_j \}$, then the new vector has
the same Euclidean norm and is still in the coset $\bfc + \complexLc$.
Therefore, the number of shortest vectors in $\bfc + \complexLc$ is
$4^{w_E^{\min}(\calC)}$.

\subsection{$\realL$ in (\ref{eq:construction-D-real-lattice}) is a Lattice}\label{app:lattice-proof}

Let $\repr{\bfg}_j = \emb{\sigma}(\bfg_j)$, for $j = 1, \ldots, k_s$.
It is easy to check that $\bflambda \in \realL$ if and only if
$\bflambda = p^s \bfr + \sum_{j = 1}^{k_s} c_{j} \repr{\bfg}_j$ for some
$\bfr \in \ZZ^n$ and $c_j \in \{0, \ldots, p^s - 1\}$ satisfying the
division condition: when $k_t< j \le k_{t+1}$, $p^t \mid c_j$ (where $t
= 1, \ldots, s-1$).

Let $\bflambda_i = p^s \bfr_i + \sum_{j = 1}^{k_s} c_{ij} \repr{\bfg}_j$
($i = 1, 2$) be two vectors from $\realL$. Then we have $\bfr_1, \bfr_2
\in \ZZ^n$, and $c_{1j}, c_{2j} \in \{0, \ldots, p^s - 1\}$ satisfy the
division condition. Now consider the difference
\[
	\bflambda_1 - \bflambda_2 = p^s (\bfr_1 - \bfr_2)
	 + \sum_{j = 1}^{k_s} (c_{1j} - c_{2j}) \repr{\bfg}_j.
\]
We will show that $\bflambda_1 - \bflambda_2 \in \realL$. We need the
following lemma from elementary arithmetic.
\begin{lem}
Let $a, d \in \ZZ$ with $d \ne 0$. Then there exist unique $q, r \in
\ZZ$ such that $a = qd + r$ and $0 \le r < |d|$.
\end{lem}

Using the above lemma, we have $c_{1j} - c_{2j} = q_j p^s + r_j$ for
some $q_j \in \ZZ$ and $r_j \in \{ 0, \ldots, p^s - 1 \}$. Furthermore,
if $p^t$ divides $c_{1j} - c_{2j}$, then $p^t$ divides $r_j$, where $t =
1, \ldots, s-1$. Thus, $\{ r_j \}$ satisfy the division condition. Note
that
\[
	\bflambda_1 - \bflambda_2 = p^s (\bfr_1 - \bfr_2 +
	\sum_{j} q_j \repr{\bfg}_j) + \sum_{j} r_j \repr{\bfg}_j.
\]
Thus, $\bflambda_1 - \bflambda_2 \in \realL$, which implies that
$\realL$ is indeed a lattice.

Next, we will construct a generator matrix for $\realL$. Let
$\repr{\bfG}$ denote the matrix with rows $\repr{\bfg}_1, \ldots,
\repr{\bfg}_n$. Clearly, we have $\det(\repr{\bfG}) = 1$ due to the way
$\{ \bfg_i \}$ are constructed. This implies that $\repr{\bfg}_1,
\ldots, \repr{\bfg}_n$ span $\ZZ^n$ over $\ZZ$. That is, any vector
$\bfr \in \ZZ^n$ can be expressed as an integer combination of
$\repr{\bfg}_1, \ldots, \repr{\bfg}_n$. Consider the set of all integer
combinations of the following vectors: $\repr{\bfg}_1, \ldots,
\repr{\bfg}_{k_1}$, $p\repr{\bfg}_{k_1 + 1}, \ldots, p\repr{\bfg}_{k_2}$,
$\ldots$, $p^s \repr{\bfg}_{k_s + 1}, \ldots, p^s \repr{\bfg}_{n}$. On
the one hand, it is easy to see that any integer combination of these
vectors is a lattice point in $\realL$. On the other hand, let
$\bflambda = p^s \bfr + \sum_{j = 1}^{k_s} c_{j} \repr{\bfg}_j$ be a
lattice point in $\realL$, where $\bfr \in \ZZ^n$ and $\{ c_j \}$
satisfy the division condition. Recall that $\bfr = \sum_{j = 1}^n b_j
\repr{\bfg}_j$ for some $b_j \in \ZZ$. Thus, we have
\[
	\bflambda = \sum_{i = 1}^{k_s} (c_i + p^s b_i) \repr{\bfg}_i
	+ \sum_{j = k_s + 1}^{n} p^s b_j \repr{\bfg}_j.
\]
Since $p^t \mid c_i$, when $k_t < i \le k_{t+1}$, we have $p^t \mid c_i
+ p^t b_i$, when $k_t < i \le k_{t+1}$. Hence, $\bflambda$ is indeed an
integer combination of the above vectors. Let $\bfG_{\realL}$ be the
matrix formed by these vectors. Then $\bfG_{\realL}$ is a generator
matrix for $\realL$.

\subsection{Proof of Relation (\ref{eq:lift_d})}\label{app:relation}

The following two observations simplify the proof of the relation
(\ref{eq:lift_d}). First, it suffices to consider the case of $s = 2$,
since the case of $s > 2$ is essentially the same. Second, it suffices
to prove the relation for the pair of nested $\ZZ$-lattices $\realL
\supseteq \realLc$, i.e.,
\begin{equation}\label{eq:relation}
\bfG_{\realLc} = {\rm diag}(\underbrace{p^{2}, \ldots, p^{2}}_{k_1},
\underbrace{p, \ldots, p}_{k_2 - k_1}, \underbrace{1, \ldots, 1}_{n -
k_2}) \bfG_{\realL}
\end{equation}
due to the lifting operation.

Next we will construct two generator matrices $\bfG_{\realL}$ and
$\bfG_{\realLc}$ satisfying the above relation. Let $\repr{\bfg}_i$
denote $\emb{\sigma}(\bfg_i)$, for $i = 1, \ldots, n$. On the one hand,
by Appendix~\ref{app:lattice-proof}, there exists a generator matrix
$\bfG_{\realL}$ of $\realL$ consisting of basis vectors $\repr{\bfg}_1,
\ldots, \repr{\bfg}_{k_1}$, $p\repr{\bfg}_{k_1 + 1}, \ldots,
p\repr{\bfg}_{k_2}$, $p^2 \repr{\bfg}_{k_2 + 1}, \ldots, p^2
\repr{\bfg}_{n}$. On the other hand, the vectors $\{ p^2 \repr{\bfg}_1,
\ldots, p^2 \repr{\bfg}_n \}$ form a basis of $\realLc$, because
$\repr{\bfg}_1, \ldots, \repr{\bfg}_n$ span $\ZZ^n$ over $\ZZ$. By
comparing these two bases for $\realL$ and $\realLc$, we conclude that
there exist two generator matrices $\bfG_{\realL}$ and $\bfG_{\realLc}$
satisfying Relation (\ref{eq:relation}).

\subsection{Proof of Proposition~\ref{prop:construction_d}}\label{app:construction-d}

It suffices to consider the case $s = 2$, since the case of $s > 2$ is
essentially the same. Consider a lattice point $\bflambda \in \realL
\setminus \realLc$ given by
\[
	\bflambda = p^2 \bfr + \sum_{j = 1}^{k_1} \beta_{1j} \repr{\bfg}_j + \sum_{j = 1}^{k_2} p \beta_{2j} \repr{\bfg}_j,
\]
where $\beta_{ij} \in \{ 0, \ldots, p - 1 \}$. Clearly, some
$\beta_{ij}$ must be nonzero, because otherwise $\bflambda = p^2 \bfr
\in \realLc$. We consider the following two cases.

Case 1: some $\beta_{1j}$ is nonzero. In this case, we construct a new
lattice ${\realL}_1 = \{ p \bfr + \sum_{j = 1}^{k_1} \beta_{j}
\repr{\bfg}_j : \bfr \in \ZZ^n, \beta_j \in \{0, \ldots, p-1\} \}$ and a
new sublattice ${\realL}_1' = \{ p \bfr: \bfr \in \ZZ^n \}$. Clearly,
we have $\bflambda \in {\realL}_1$ and $\bflambda \notin {\realL}_1'$.
Thus, $\bflambda \in {\realL}_1 \setminus {\realL}_1'$. Note that the
nested lattice pair ${\realL}_1 \supseteq {\realL}_1'$ can be obtained
from the code $\calC_1$ by Construction~A. Thus, we have $\| \bflambda
\|^2 \ge w_E^{\min}(\calC_1)$ and the number of lattice points
$\bflambda$ of the Euclidean weight $w_E^{\min}(\calC_1)$ is upper
bounded by $K({\realL}_1 / {\realL}_1')$.

Case 2: all $\beta_{1j}$ are zero, and some $\beta_{2j}$ is nonzero. In
this case, we construct a new lattice ${\realL}_2 = \{ p \bfr + \sum_{j
= 1}^{k_2} \beta_{j} \repr{\bfg}_j : \bfr \in \ZZ^n, \beta_j \in \{ 0,
\ldots, p-1 \} \}$ and a new sublattice ${\realL}'_2 = \{ p \bfr: \bfr
\in \ZZ^n \}$. Clearly, we have $\bflambda = p^2 \bfr + \sum_{j =
1}^{k_2} p \beta_{2j} \repr{\bfg}_j \in p {\realL}_2$ and $\bflambda
\notin p {\realL}_2'$. Thus, $\bflambda \in p {\realL}_2 \setminus p
{\realL}_2'$. Similar to Case~1, the nested lattice pair ${\realL}_2
\supseteq {\realL}_2'$ can be obtained from the code $\calC_2$ by
Construction~A. Thus, we have $\| \bflambda \|^2 \ge p^2
w_E^{\min}(\calC_2)$, and the number of lattice points $\bflambda$ of
the Euclidean weight $w_E^{\min}(\calC_2)$ is upper bounded by
$K({\realL}_2 / {\realL}_2')$.

Combining the above two cases, we have,
for all $\bflambda \in
\realL \setminus \realLc$,
that
$\| \bflambda \|^2 \ge \min\{
w_E^{\min}(\calC_1), p^2 w_E^{\min}(\calC_2) \}$,
which implies that $d^2(\realL / \realLc) \ge
\min\{ w_E^{\min}(\calC_1), p^2 w_E^{\min}(\calC_2) \}$. Recall that
$\complexL = \realL \times \realL$. Hence, we have
\begin{align*}
	d^2(\complexL / \complexLc) &= d^2(\realL / \realLc) \\
	&\ge \min\{ w_E^{\min}(\calC_1), p^2 w_E^{\min}(\calC_2) \}.
\end{align*}
Note that $V(\complexLc) = p^{4n}$ and $V(\complexLc) / V(\complexL) =
p^{2(k_1 + k_2)}$, since each $\beta_{ij} \in \{ 0, \ldots, p-1 \}$.
Hence, we have $V(\complexL) = p^{2(2n - k_1 - k_2)}$ and
\begin{align*}
	\gamma_c(\complexL / \complexLc) &= d^2(\complexL /\complexLc)/p^{2(2 - (k_1 + k_2)/n )} \\
	&\ge \frac{\min\{ w_E^{\min}(\calC_1),
p^2 w_E^{\min}(\calC_2) \}}{p^{2(2 - (k_1 + k_2)/n )}}.
\end{align*}

We also have $K(\realL / \realLc) \le K({\realL}_1 / {\realL}_1')
+ K({\realL}_2 / {\realL}_2')$ and $K(\complexL / \complexLc) =
2K(\realL / \realLc)$, completing the proof for the case $s = 2$.

\subsection{Modified Viterbi Decoder for Example~\ref{exam:code-construction}}\label{app:Viterbi}

We will show that the nearest neighbor quantizer
$\calQ_{\complexL}^{\nn}$ can be implemented through a modified Viterbi
decoder.

First, note that $\calQ_{\complexL}^{\nn}$ solves the following
optimization problem
\begin{align}
		\mbox{minimize} &\quad \| \bflambda - \alpha \bfy \| \label{eq:opt1} \\
		\mbox{subject to} &\quad \bflambda \in \complexL. \nonumber
\end{align}

Second, note that the problem~(\ref{eq:opt1}) is equivalent to
\begin{align}
		\mbox{minimize} &\quad \| \emb{\sigma}(\bfc) + \bflambda'
		 - \alpha \bfy \| \label{eq:opt2} \\
		\mbox{subject to} &\quad \bfc \in \calC  \\
		&\quad \bflambda' \in \complexLc . \nonumber
\end{align}
This is because each lattice point $\bflambda \in \complexL$ can be
expressed as $\bflambda = \emb{\sigma}(\bfc) + \complexLc$, where $\bfc =
\sigma(\bflambda)$ and $\bflambda' \in \complexLc$.

Third, note that Problem~(\ref{eq:opt2}) is equivalent to
\begin{align}
\mbox{minimize} &\quad \| [\emb{\sigma}(\bfc)
- \alpha \bfy] \bmod \complexLc \| \label{eq:opt3} \\
\mbox{subject to} &\quad \bfc \in \calC, \nonumber
\end{align}
where $[\bfx] \bmod \complexLc$ is defined as $[\bfx] \bmod \complexLc
\triangleq \bfx - \calQ_{\complexLc}^{\nn}(\bfx)$. This is because
$\bflambda' = - \calQ_{\complexLc}^{\nn}(\emb{\sigma}(\bfc)
- \alpha \bfy)$ solves Problem~(\ref{eq:opt2}) for any $\bfc \in \calC$.

Now it is easy to see the problem~(\ref{eq:opt3}) can be solved through
a modified Viterbi decoder with the metric given by $\|[\cdot] \bmod
\complexLc\|$ instead of $\|\cdot \|$. Therefore, the nearest neighbor
quantizer $\calQ_{\complexL}^{\nn}$ can be implemented through a
modified Viterbi decoder.

\subsection{Proof of Theorem~\ref{thm:strongly}}\label{app:dominant}

First, we show the existence of the solution $\{ \bfa_1, \ldots, \bfa_m
\}$ by induction on $m$.

If $m = 1$, then the vector $\bfa_1$ can be chosen such that $\bfa_1 \bfL$
is one of the shortest lattice points. Note that $\bfa_1$ is not
divisible by $\pi$; otherwise it will not be one of the shortest lattice
points. In other words, $\proj{\bfa}_1$ is indeed nonzero. Hence, the
solution $\bfa_1$ always exists when $m = 1$.

Now suppose the solution $\{ \bfa_1, \ldots, \bfa_k \}$ exists when $k <
m$. We will show the existence of the vector $\bfa_{k+1}$.

Consider the following set
\[
\calA = \{ \bfa \in T^L : \proj{\bfa}_1, \ldots, \proj{\bfa}_{k}, \proj{\bfa} \ \text{are linearly independent} \}.
\]
Clearly, the set $\calA$ is nonempty, since $k < m$. Then the vector
$\bfa_{k+1}$ can be chosen as
\[
\bfa_{k+1} = \arg\min_{\bfa \in \calA} \|\bfa \bfL \|.
\]
This proves the existence of the vector $\bfa_{k + 1}$, which completes
the induction.

Second, we show that the solution $\{ \bfa_1, \ldots, \bfa_m \}$ is a
dominant solution by induction on $m$.

If $m = 1$, then $\|\bfa_1 \bfL\| \le \| \bfb_1 \bfL \|$ for any feasible
solution $\bfb_1$, since $\bfa_1 \bfL$ is one of the shortest lattice
points.

Now suppose that $\{ \bfa_1, \ldots, \bfa_k \}$ is a dominant solution
when $k < m$. We will show that $\{ \bfa_1, \ldots, \bfa_k, \bfa_{k+1} \}$
is also a dominant solution.

Suppose that $\{ \bfb_1, \ldots, \bfb_k, \bfb_{k+1} \}$ is a feasible solution
with $\| \bfb_1 \bfL\| \le \ldots \le \| \bfb_{k+1} \bfL\|$. Since
$\proj{\bfb}_1, \ldots, \proj{\bfb}_k$ are linearly independent, we have
\[
\| \bfa_i \bfL \| \le \| \bfb_i \bfL \|, \ i = 1, \ldots, k.
\]
It remains to show $\| \bfa_{k+1} \bfL \| \le \| \bfb_{k+1} \bfL \|$. We consider
the following two cases.

\begin{enumerate}
\item If there exists some $\bfb_i$ ($i = 1, \ldots, k+1$) such that
$\proj{\bfa}_1, \ldots, \proj{\bfa}_k,\proj{\bfb}_i$ are linearly
independent, then by the construction of $\bfa_{k+1}$, we have
\[
\| \bfa_{k+1} \bfL \| \le \| \bfb_i \bfL \| \le \| \bfb_{k+1} \bfL \|.
\]
\item Otherwise, each $\proj{\bfb}_i$ can be expressed as a linear
combination of $\proj{\bfa}_1, \ldots, \proj{\bfa}_k$. That is,
\[
\proj{\bfb}_i \in \mbox{Span}\{ \proj{\bfa}_1, \ldots, \proj{\bfa}_k\}.
\]
This is contrary to the fact that $\proj{\bfb}_1, \ldots,
\proj{\bfb}_{k+1}$ are linearly independent, since any $k+1$ vectors in a
vector space of dimension $k$ are linearly dependent.
\end{enumerate}

Therefore, we have $\| \bfa_{k+1} \bfL \| \le \| \bfb_{k+1} \bfL \|$, which
completes the induction.


\begin{IEEEbiographynophoto}{Chen Feng}
received the B.E. degree from Shanghai Jiao Tong University in 2006
and the M.A.Sc.\ degree from the University of Toronto in 2009.
He is currently a
Ph.D. student in the Department of Electrical and Computer Engineering,
University of Toronto. His research interests are in network coding, coding
theory, information theory, and their applications to computer networking.

During his Ph.D. studies, Chen Feng won several awards for his academic
achievement, including the Chinese Government Award for Outstanding Students
Abroad in 2012, and the Shahid U. H. Qureshi Memorial Scholarship in 2013.
\end{IEEEbiographynophoto}

\begin{IEEEbiographynophoto}{Danilo Silva}
received the B.Sc.\ degree from the Federal University of Pernambuco (UFPE),
Recife, Brazil, in 2002, the M.Sc.\ degree from
the Pontifical Catholic University of Rio de Janeiro (PUC-Rio), Rio de Janeiro,
Brazil, in 2005, and the Ph.D. degree from the University of Toronto, Toronto,
Canada, in 2009, all in electrical engineering.

From 2009 to 2010, he was a Postdoctoral Fellow at the University of Toronto,
at the \'Ecole Polytechnique F\'ed\'erale de
Lausanne (EPFL), and at the State University of Campinas (UNICAMP).
In 2010, he joined the Department of Electrical Engineering,
Federal University of Santa Catarina (UFSC), Brazil, where
he is currently an Assistant Professor. His research interests include channel
coding, information theory, and network coding.

Dr. Silva was a recipient of a CAPES Ph.D. Scholarship in 2005, the
Shahid U. H. Qureshi Memorial Scholarship in 2009, and a FAPESP
Postdoctoral Scholarship in 2010.
\end{IEEEbiographynophoto}

\begin{IEEEbiographynophoto}{Frank R. Kschischang}
received the B.A.Sc.\ degree (with honors) from the University of British
Columbia in 1985 and the M.A.Sc.\ and Ph.D.\ degrees from the University of
Toronto in 1988 and 1991, respectively, all in electrical engineering.  He
is a Professor and Canada Research Chair at the University of Toronto,
where he has been a faculty member since 1991.  Between 2011 and 2013 he
was a Hans Fischer Senior Fellow at the Institute for Advanced Study,
Technische Universit\"at  M\"unchen.

His research interests are focused primarily on the area of channel coding
techniques, applied to wireline, wireless and optical communication systems
and networks.  He is the recipient of the 2010 Killam Research Fellowship,
the 2010 Communications Society and Information Theory Society Joint Paper
Award and the 2012 Canadian Award in Telecommunications Research.  He is a
Fellow of IEEE, of the Engineering Institute of Canada, and of the Royal
Society of Canada.

During 1997-2000, he served as an Associate Editor for Coding Theory for
the \textsc{IEEE Transactions on Information Theory}.  He also served as
technical program co-chair for the 2004 IEEE International Symposium on
Information Theory (ISIT), Chicago, and as general co-chair for ISIT 2008,
Toronto.  He served as the 2010 President of the IEEE Information Theory
Society.
\end{IEEEbiographynophoto}

\end{document}